\documentclass[twocolumn]{aastex63}
\usepackage{enumerate}
\usepackage{multirow}
\usepackage{graphicx}
\usepackage{lineno}
\usepackage[normalem]{ulem}
\usepackage{xcolor}
\usepackage{hyperref}
\usepackage{tabularx}
\usepackage{verbatim}
\usepackage{booktabs}
\usepackage{afterpage}
\hypersetup{linkcolor=blue,citecolor=blue,filecolor=cyan,urlcolor=blue}
\usepackage{amsmath}
\usepackage{kotex}

\newcommand{\vect}[1]{\boldsymbol{#1}}
\def\hMsun{\,h^{-1}M_{\odot}}

\def\Mpc{\,h^{-1}{\rm {Mpc}}}
\def\kpc{\,h^{-1}{\rm {kpc}}}

\def\angsl{\theta_{\rm SL}}

\def\cossl{\cos\theta_{\rm SL}}

\def\subfigc#1#2{\hfill\vbox{\parskip=0pt\hsize=#2
\includegraphics[width=#2]{#1}\vskip-6pt}\hfill}

\revised{\today}

\shorttitle{Dissecting the Spin--Orbit Alignment of Halos}
\shortauthors{An et al.}

\begin{document}

\title{Living with Neighbors. IV. Dissecting the Spin$-$Orbit Alignment of Dark Matter Halos: \\Interacting Neighbors and the Local Large-scale Structure}

\correspondingauthor{Suk-Jin Yoon}
\email{sjyoon0691@yonsei.ac.kr}

\author[0000-0003-3791-0860]{Sung-Ho An}
\affiliation{Department of Astronomy, Yonsei University, Seoul 03722, Republic of Korea}
\affiliation{Center for Galaxy Evolution Research, Yonsei University, Seoul 03722, Republic of Korea}

\author[0000-0002-4391-2275]{Juhan Kim}
\affiliation{Center for Advanced Computation, Korea Institute for Advanced Study, Seoul 02455, Republic of Korea}

\author[0000-0001-7075-4156]{Jun-Sung Moon}
\affiliation{Department of Astronomy, Yonsei University, Seoul 03722, Republic of Korea}
\affiliation{Center for Galaxy Evolution Research, Yonsei University, Seoul 03722, Republic of Korea}

\author[0000-0002-1842-4325]{Suk-Jin Yoon}
\affiliation{Department of Astronomy, Yonsei University, Seoul 03722, Republic of Korea}
\affiliation{Center for Galaxy Evolution Research, Yonsei University, Seoul 03722, Republic of Korea}

\begin{abstract}

Spin--orbit alignment (SOA; i.e., the vector alignment between the halo spin and the orbital angular momentum of neighboring halos) provides an important clue to how galactic angular momenta develop.
For this study, we extract virial-radius-wise contact halo pairs with mass ratios between 1/10 and 10 from a set of cosmological $N$-body simulations.
In the spin--orbit angle distribution, we find a significant SOA in that 52.7\,\%\,$\pm$\,0.2\,\% of neighbors are on the prograde orbit.
The SOA of our sample is mainly driven by low-mass target halos ($<10^{11.5}\hMsun$) with close merging neighbors, corroborating the notion that the tidal interaction is one of the physical origins of SOA.
We also examine the correlation of SOA with the adjacent filament and find that halos closer to the filament show stronger SOA.
Most interestingly, we discover for the first time that halos with the spin parallel to the filament experience most frequently the prograde-polar interaction (i.e., fairly perpendicular but still prograde interaction; spin--orbit angle $\sim$ 70$^{\circ}$).
This instantly invokes the spin-flip event and the prograde-polar interaction will soon flip the spin of the halo to align it with the neighbor's orbital angular momentum.
We propose that the SOA originates from the local cosmic flow along the anisotropic large-scale structure, especially that along the filament, and grows further by interactions with neighbors.

\end{abstract}


\vspace{0.5cm}
\section{Introduction} \label{sec:intro}

\defcitealias{Moon19}{Paper \textrm{I}}
\defcitealias{An19}{Paper \textrm{II}}
\defcitealias{Moon21}{Paper \textrm{III}}

Recent observations have revealed that the halo spin is associated with the large-scale structure (LSS).
According to the tidal torque theory \citep{Peeb69,Whit84,Vitv02,Peir04,Bett07,Bett10,Stew13,Zava16,Zjup17}, the spin--LSS alignment can be explained by the tidal torque of the ambient anisotropic matter distribution giving rise to galaxy rotation.
Numerical simulations have shown that, for instance, less (more) massive halos in filaments tend to have a spin parallel (perpendicular) to the filament's spine \citep{Arag07,Hahn07,Codi15,Laig15,WK17}.
Galaxy observations also confirm this spin--LSS alignment (\citealt{Temp13,BB20,Welk20}; see also \citealt{Krol19}).
On the other hand, recent studies have reported a possible link between the spin (and the shape) of centrals and the spatial distribution of their satellites.
Such spin--satellite alignment is found among dark matter halos \citep{Wang14,Kang15} and among galaxies \citep{Brai05,Yang06,Agus10,Dong14,Temp15,Vell15b,Shao16,Wang18sca}.
The alignment depends on the mass \citep{Vell15b}, morphology \citep{Wang10}, and color of the central galaxies \citep{Wang18sca} and on the color of the satellites \citep{Dong14}.
The spin--satellite alignment indicates that the galaxy (or halo) spin is regulated by tidal interaction.
Taking the two types of alignments together, it has been suggested that the angular momentum of a galaxy evolves with nonlinear events such as mergers and accretions along the local LSS \citep{Porc02,Vitv02,Peir04,Hetz06,Hahn07,Cerv10,Stew13,Rodr17,PR20}.

Motivated by this concept of spin--LSS and spin--satellite alignments, we explore the alignment between the spin of a halo and the orbital angular momentum of its neighbor, which we refer to as the spin--orbit alignment (SOA).
Our working hypotheses are (a) that the neighbor's orbital angular momentum can be converted into the internal angular momentum of the central galaxy \citep[e.g.,][]{Aube04,Bail05}, and (b) that interacting neighbors preferentially coming along the LSS \citep{WK18} have orbital angular momenta reflecting the local matter flow.
We expect that dark matter halo pairs will display SOA, and then the halo spin will get faster within a strong tidal field \citep{Wang11} such as in a pair system (\citealt{John19}; see also \citealt{Cerv10}).
Previous studies have found corotating satellites with their host clusters (in simulations; \citealt{Warn06}) and with the isolated galaxies hosting them (in observations; \citealt{Herb08}; but see also \citealt{Hwan10}).
\citet[][hereafter \citetalias{Moon21}]{Moon21} proposed that SOA can be developed by interactions with neighboring galaxies.
Some studies have found spin--spin alignment in interacting pairs \citep{Mesa14,Koo18} although it is still controversial \citep{Cerv10,Buxt12,LeeJ12}.
This work focuses on SOA for dark matter halos with virial-radius-wise interacting neighbors using a set of cosmological $N$-body simulations.
We intend to address the question of which physical parameter dominantly influences the halo spin.
To this end, we measure the SOA and examine its dependence on the halo mass, pairwise distance, interaction type, large-scale density, distances from the nearest filament and node, and the angle between the halo spin vector and the filament axis.

The present series of papers investigates both observationally and theoretically the impact of interacting neighbor galaxies (or halos) on galactic properties. 
\citeauthor{Moon19} (\citeyear{Moon19}, \citetalias{Moon19}) revealed how strongly the nearest neighbor affects the star formation activity of galaxies using a comprehensive galaxy pair catalog from the Sloan Digital Sky Survey \citep{York00}. 
\citeauthor{An19} (\citeyear{An19}, hereafter \citetalias{An19}), using a set of cosmological $N$-body simulations, found that flyby-type interactions substantially outnumber merger-type interactions toward $z=0$, and that the flyby contributes to the galactic evolution more significantly than ever at the present epoch.
\citetalias{Moon21} found a strong SOA of close galaxy pairs using IllustrisTNG simulation data \citep{Mari18, Naim18, Nels18, Pill18, Spri18}, and the results imply that the interacting neighbor compels the spin vector of its target to be aligned with its orbital angular momentum.
This fourth paper is an extension of \citetalias{An19} and \citetalias{Moon21}.
By extending the simulation dataset of \citetalias{An19}, we aim to understand the buildup process of galactic angular momentum via the SOA on a larger scale of a few hundred kiloparsecs compared to $\lesssim 100 \kpc$ explored by \citetalias{Moon21}.
Moreover, this paper is the first of its kind to quantitatively investigate the association of the SOA with the LSS.

Throughout the paper, we assume that the dark matter halo spin represents the galaxy spin.
The spin vector (or minor axis) of dark matter halos has been, however, reported to be somewhat misaligned with the spin vector (or minor axis) of baryonic components (so-called galaxy--halo misalignment).
The degree of the misalignment depends on the halo and galaxy properties \citep{Bail05b,Vell15a,Gane19}.
Despite the presence of the galaxy--halo misalignment, the mean angle offset between the two components' vectors ($25^{\circ}$--$50^{\circ}$; \citealt{Bail05,Shao16,Chis17}) is smaller than the offset ($57^{\circ}$) expected from the random distribution.
We further suppose that the tidal interaction eventually adjusts the halo spin to align with the galaxy spin because tidal interaction affects the spin orientation \citep{Hetz06,Cerv10,Wang11,John19}.
This implies that the SOA makes the present galaxy--halo alignment stronger, an effect similar to that of satellite galaxies on the galaxy--halo alignment of their central galaxies \citep{Shao16}.
To be more conservative, we measure the halo spin vector only in the central region of the target halo.
Many simulation studies have found that a galaxy is more closely associated with the inner part of its host halo than with the outer halo \citep{Bail05b,Bett10,Shao16}.
Therefore, we use the SOA for dark matter halos as a representative of the SOA for galaxies.

This paper is organized as follows.
Section 2 describes our simulations, the halo and pair sample selection, and the measurement method of the SOA. 
Section 3 shows the spin--orbit angle distributions and estimates their dependence on various parameters.
In Section 4 we discuss the physical causes of SOA, linking it to both interactions with neighbors and the LSS.
We summarize our results in Section 5.

\section{Data and Analysis} \label{sec:simul}

\subsection{Halo Pair Sample and Subsamples}

We perform a series of the cosmological $N$-body simulations using the parallel tree particle-mesh code, \texttt{GOTPM} \citep[Grid-of-Oct-Tree-Particle-Mesh;][]{Dubi04}. 
There are 14 simulations in total, each of which has the same mass resolution of $M_{\rm p}$\,=\,$1.55\times10^8\hMsun$, but 10 simulations have a box with a side length of $L_\mathrm{box} = 64\Mpc$ while 4 simulations have $L_\mathrm{box} = 128\Mpc$.
All simulations are based on the WMAP 9-year cosmology \citep{Benn13} but apply different sets of random numbers to take into account the cosmic variance (see \citetalias{An19} for details).

We employ the \texttt{ROCKSTAR} \citep[Robust Overdensity Calculation using K-Space Topologically Adaptive Refinement;][]{Behr13} halo-finding algorithm to build a halo catalog.
In the 14 simulations, we choose the target halos with a halo mass range of $10^{10.8}\hMsun$ to $10^{13.0}\hMsun$ (a total of 585,739 halos at $z$\,=\,0) and identify the halos with interacting neighbors using the same criteria fully described in \citetalias{An19} except for the mass ratio.
In particular, the mass ratio range is 1/10 to 10 (while we adopted 1/3--3 in \citetalias{An19}), and the distance ($d_{\rm tn}$) between the target halo (hereafter, we use a subscript `$t$') and its neighbor (hereafter, we use a subscript `$n$') should be smaller than the sum of the virial radii of the two halos ($d_{\rm tn} < R_{\rm vir,t} + R_{\rm vir,n}$).
In this study, one target halo can have multiple neighbors, and we count all the neighbors with different orbital configurations.
The total number of paired neighbors is defined by
\linenomath{
\begin{equation}
N_{\rm sample} \equiv \sum_i^{N_{\rm t}} N_{ {\rm n},i}\,, \label{eq:nsamp} 
\end{equation}}
where $N_{{\rm n},i}$ is the number of neighbors that belong to the $i$th target halo.
At $z=0$, we have found $N_{\rm sample}$\,=\,282,098 interacting neighbors around $N_{\rm t}$\,=\,195,699 target halos with the mass ratio and distance conditions (on average, 1.44 interacting neighbors per one target halo).
In \citetalias[][]{An19}, we found that the number fraction of our multiple interactions is about five times higher than the observation by \citet{Darg11}, due to our distance criterion ($\sim300\kpc$) being longer than theirs ($30\,{\rm kpc}$).

Next, we break down the pair sample according to the halo mass, environment, and total energy of the pair system.
First, we use the virial mass, which is calculated according to the definition of \citet{Brya98}. 
Second, we define the environmental parameter ($\Phi_{\rm Env}$) as the percentile rank of the total mass of halos that are more massive than $10^{9.8}\hMsun$ within a comoving distance of $5\Mpc$ from the target halo.
Finally, with the total energy ($E_{\rm tn}$, the sum of the kinetic and potential energy) of the pair system, we classify the interacting neighbors into mergers ($E_{\rm tn}<\Delta E$) and flybys ($E_{\rm tn} \geq \Delta E$), where $\Delta E$ is the capture criterion used in \citet{Gned03}.
Gnedin presumed that some pairs with $0<E_{\rm tn}<\Delta E$ are merged ultimately due to the energy loss induced by the dynamical friction.

\begin{figure*}[ht!]
\centering
\gridline{\subfigc{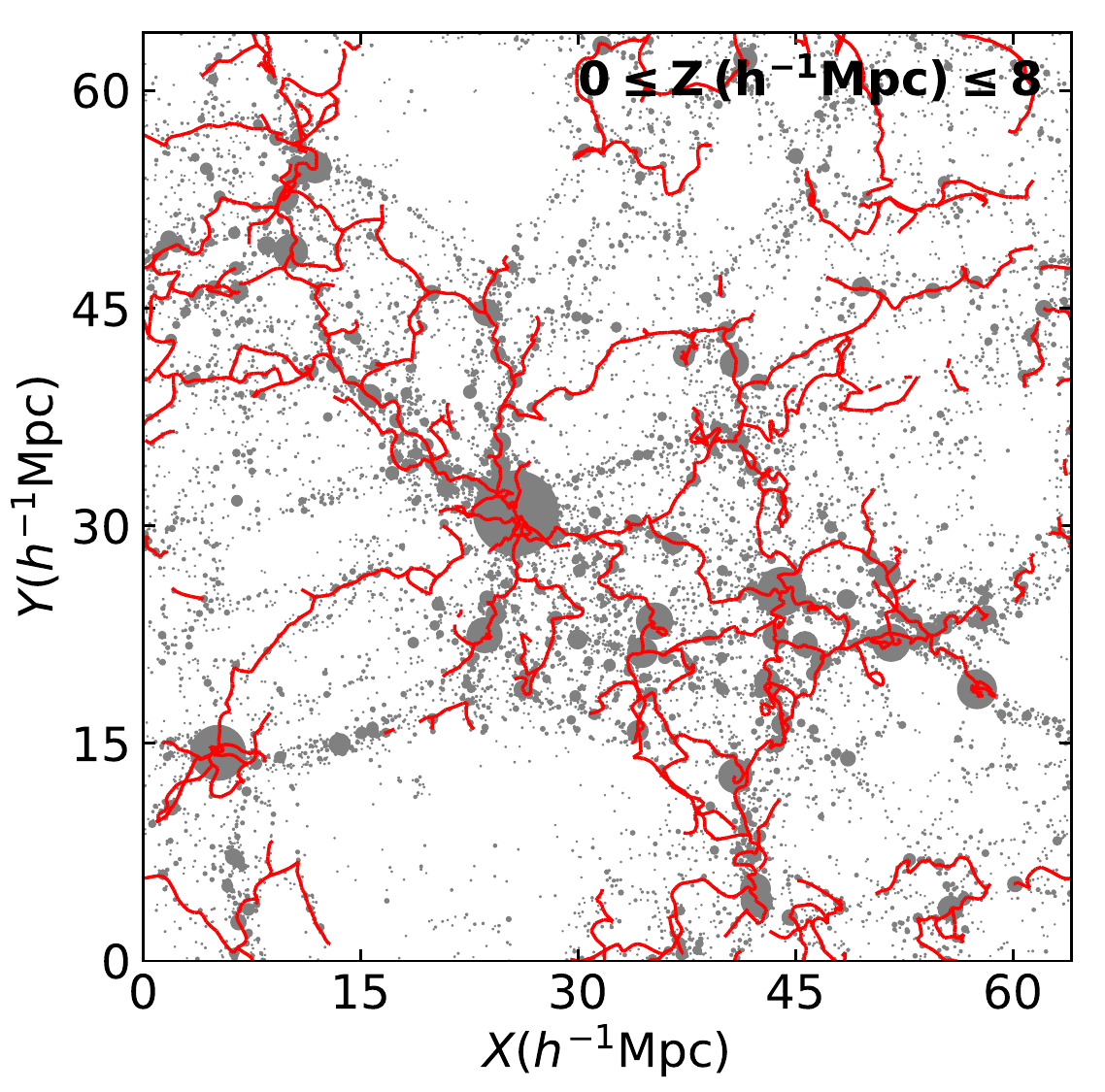}{0.33\textwidth}
\subfigc{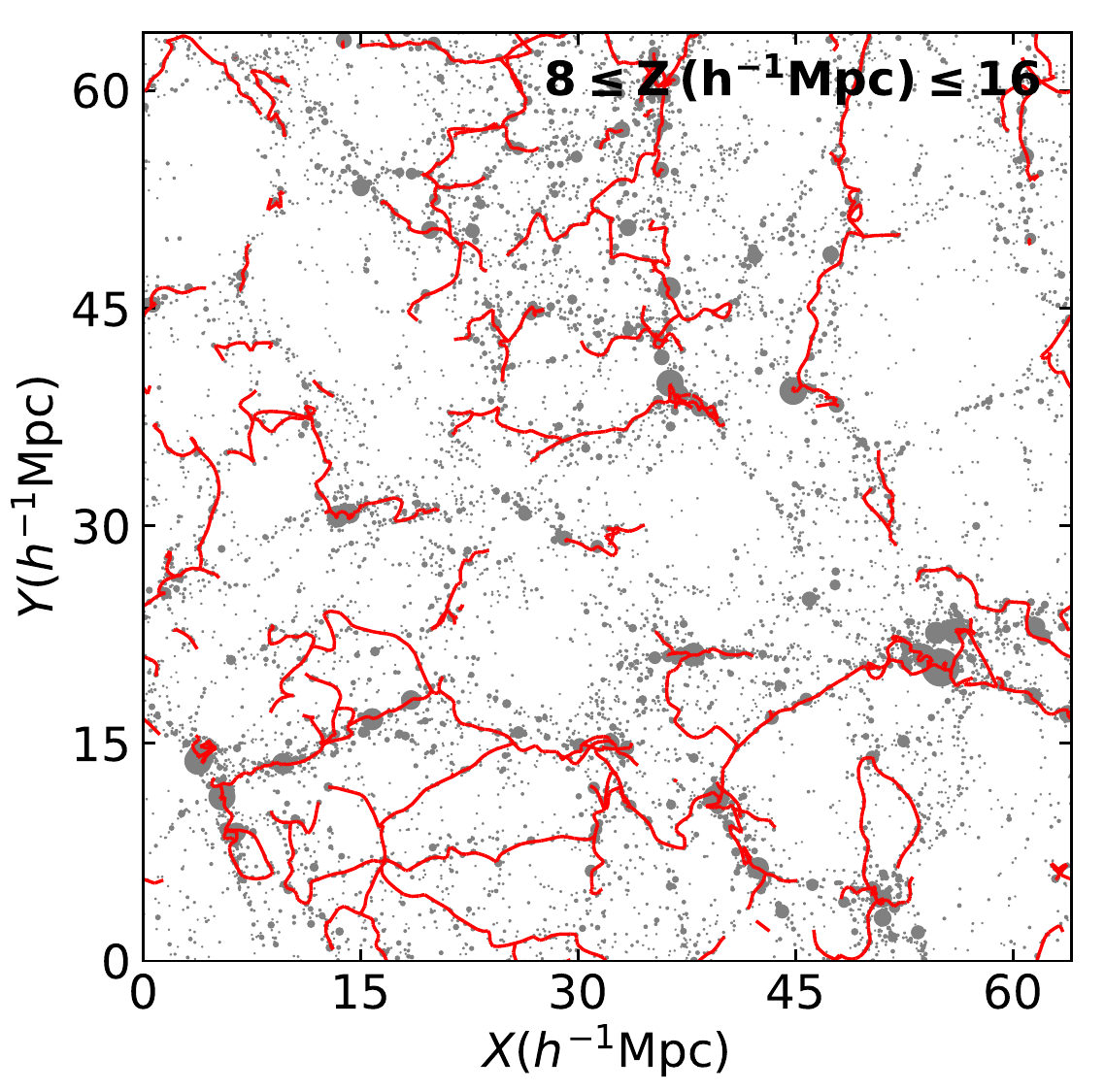}{0.33\textwidth}\subfigc{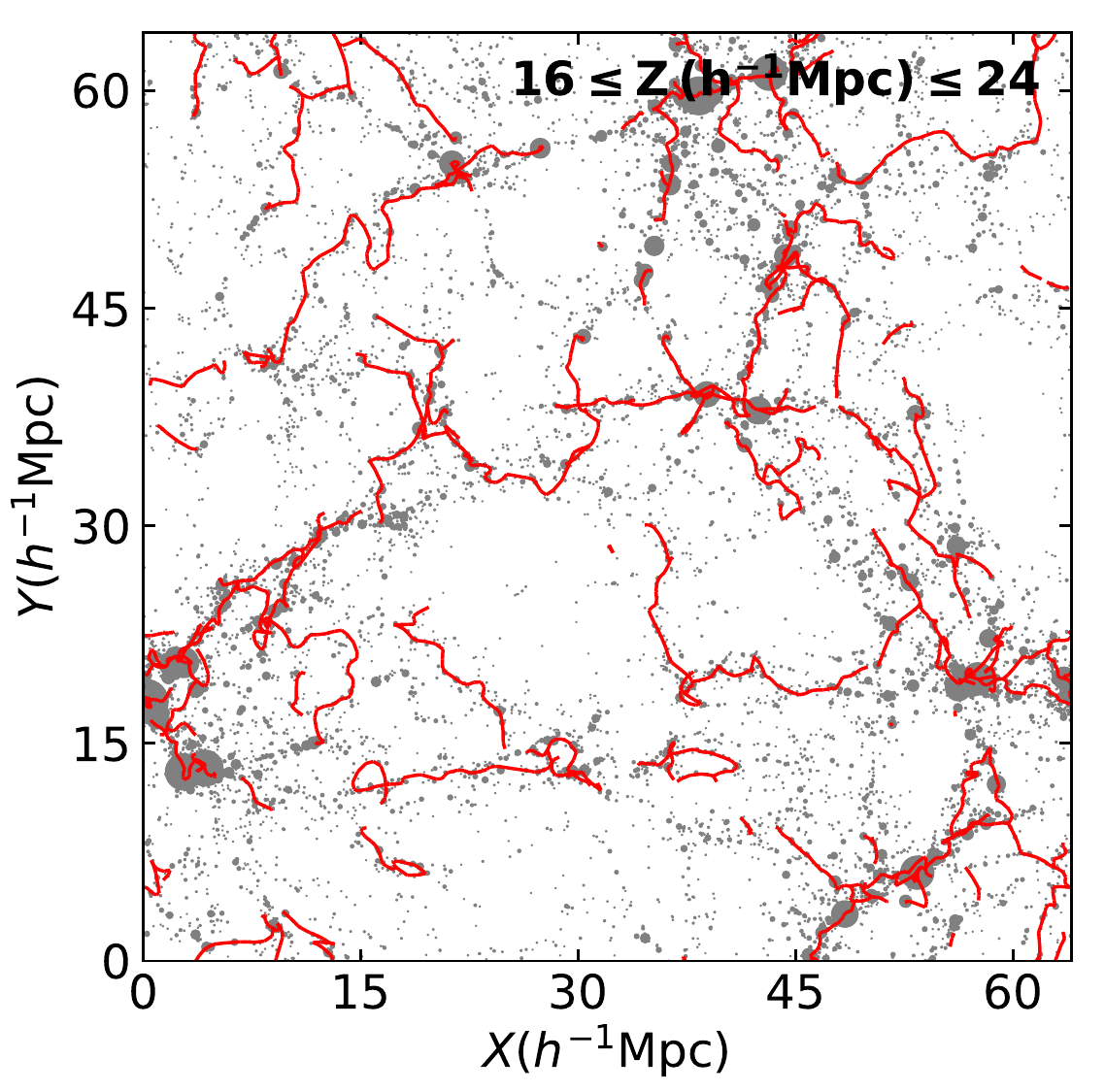}{0.33\textwidth}}
\vspace{-1ex}
\caption{Distribution of halos (gray circles) and filaments (red lines) in three back-to-back slabs with a $Z$-direction thickness of $8\Mpc$ in one of our simulations at $z=0$. The filaments are extracted from the distribution of all halos including targets and neighbors by means of \texttt{DisPerSE} \citep{Sous11a,Sous11b} with a persistence threshold of $7\sigma$. The radius of the circles indicates twice the virial radius of the halos. \label{fig:filament}}
\vspace{1.5ex}
\end{figure*}

We also measure the distances from the nearest filament and the nearest node, with respect to the center position of the target halos.
For the measurement, we extract filamentary structures from the distribution of all targets and neighbors using a publicly available code, \texttt{DisPerSE} \citep{Sous11a,Sous11b}.
From a smoothed density field constructed by a mass-weighted Delaunay tessellation, the code identifies a ridge line by linking the density maximum to another maximum passing through a saddle point.
At the same time, identified ridge lines have a persistence that represents the significance level of the lines.
In this study, we take $7\sigma$ as the proper persistence threshold to investigate the correlation of the SOA with the filaments.
If we use a lower (higher) threshold, the filamentary structures become more complicated (sparser) as shown in \citet{Codi18}, who dealt with correlations of the filament's properties with the persistence threshold.
In Figure \ref{fig:filament}, we display the distribution of halos and filaments in three consecutive slabs with a thickness of $8\Mpc$.
Seemingly discrete filaments are due to the projection effect, i.e., they are not actually broken but spread over the other slabs.

\subsection{Spin--Orbit Angle Measurements} \label{sec:soangmeasure}

To quantify the SOA, we first measure the spin--orbit angle ($\angsl$), i.e., the angle between the spin vector of a target halo ($\vect{S}$) and the orbital angular momentum vector of its neighbor ($\vect{L}$).
Prior to the vector measurement, we take into account the definition of the position and velocity of a halo, to minimize the effect of tidal stripping during contact interaction.
The halo's position and velocity are calculated by using the bound member particles in the innermost region (about 10\,\% of the halo radius; see \citealt{Behr13} for details).
We then measure the spin vector as a sum of the angular momenta of the bound member particles in the innermost region instead of the whole member particles ($\vect{S}\equiv\vect{S_{\rm c}}\not\equiv\vect{S_{\rm w}}$).
This helps us to better trace the galaxy spin \citep[e.g.,][]{Shao16}.
The vector $\vect{L}$ ($\equiv\vect{d_{\rm tn}}\times\vect{V_{\rm tn}}$) is the cross product of the center position of a neighbor halo from the target halo center ($\vect{d_{\rm tn}}$) and the relative velocity ($\vect{V_{\rm tn}}$) with consideration of the Hubble flow.

We note that the halo spin vector measurement can be contaminated by the neighbor's orbital angular momentum (see \citealt{Moon21} for details).
According to the halo finding algorithm, if two interacting halos are so close that their virial radii overlap with each other, the larger halo takes the outskirts of the smaller halo as well as its own member particles \citep[e.g.,][]{Muld11}.
Such misidentification hinders an accurate spin vector measurement.
We remedy the problem by using the core spin vector instead of the whole spin vector, but for very close pairs it is unavoidable.

\begin{figure*}[ht!]
\centering
\includegraphics[width=\textwidth]{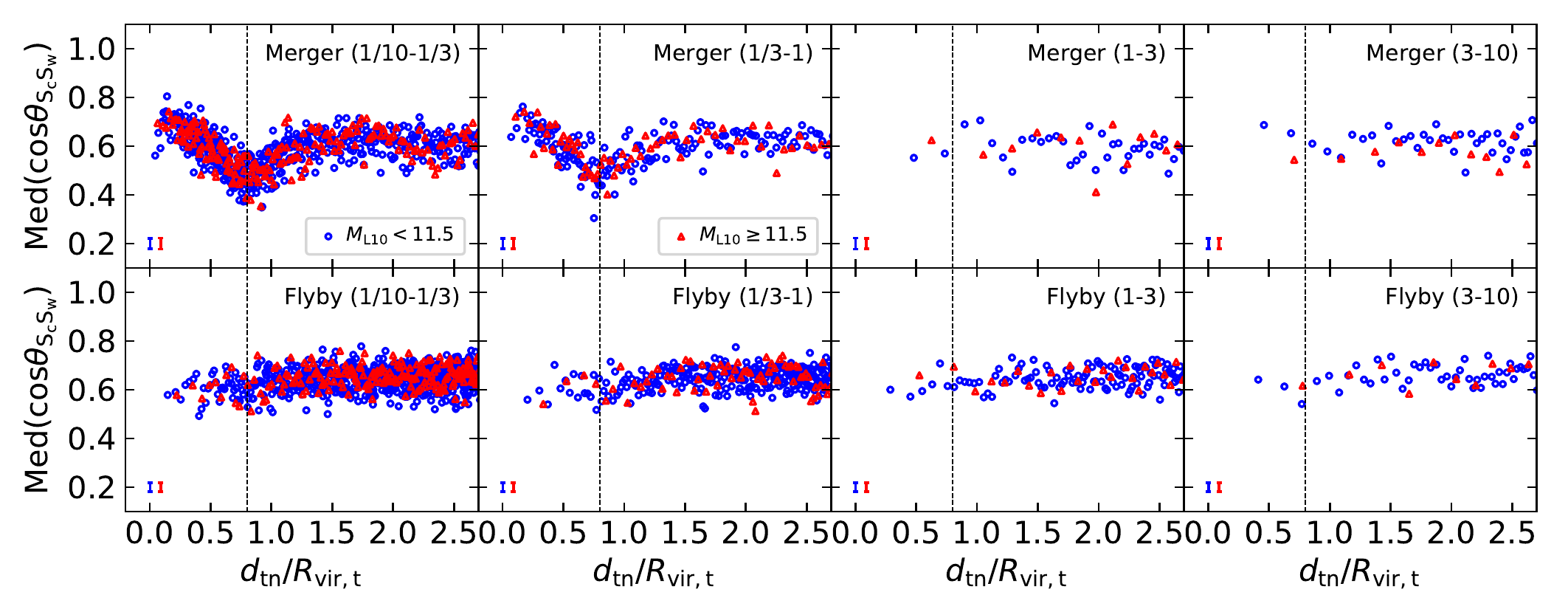}
\vspace{-5ex}
\caption{Median values of offset angles ($\cos\theta_{\rm S_{c}S_{w}}$) between the core halo spin vector ($\vect{S_{\rm c}}$) and the whole halo spin vector ($\vect{S_{\rm w}}$) as a function of the pairwise distance normalized to the virial radius of the target halos ($d_{\rm tn}/R_{\rm vir,t}$). The top panels are for the target halos with merging neighbors and the bottom panels are for those with flybying neighbors. Each column shows the distribution at a given mass ratio range: 1/10--1/3 (first column), 1/3--1 (second), 1--3 (third), and 3--10 (fourth). The pair sample is divided into two subsamples: $M_{\rm L10}<11.5$ (blue circles) and $M_{\rm L10}\geq11.5$ (red triangles). 
Each of the symbols represents the median of 200 target halos, and the standard errors are indicated in the bottom left corner of the panels.
Vertical dashed lines represent $d_{\rm tn}=0.8R_{\rm vir,t}$. \label{fig:ssoff}}
\vspace{1.5ex}
\end{figure*}

Figure \ref{fig:ssoff} shows the median cosine of offset angles ($\cos\theta_{\rm S_{\rm c}S_{\rm w}}$) between the core spin vector ($\vect{S_{\rm c}}$) and the whole spin vector ($\vect{S_{\rm w}}$) of our target halos as a function of pairwise distance normalized to the target halo's virial radius ($d_{\rm tn}/R_{\rm vir,t}$).
Overall, the median offset angle for distant pairs ($d_{\rm tn}>1.5R_{\rm vir,t}$) is almost constant regardless of the halo mass, mass ratio, and interaction type.
For close pairs ($d_{\rm tn}<1.5R_{\rm vir,t}$), the trend depends on the interaction type and mass ratio.
In cases of larger merging neighbors ($M_{\rm n}/M_{\rm t}>1$) and flybying neighbors, the median value of $\cos\theta_{\rm S_{\rm c}S_{\rm w}}$ is almost constant.
In contrast, in cases of smaller merging neighbors ($M_{\rm n}/M_{\rm t}<1$), the median offset angle decreases at $0.8R_{\rm vir,t}<d_{\rm tn}<1.5R_{\rm vir,t}$ and increases at $d_{\rm tn}<0.8R_{\rm vir,t}$ with decreasing distance---that is, as the neighbors get close to their targets, the inaccurate assignment of the outskirts of the neighbors results in the erroneous measurement of the whole spin vector ($0.8R_{\rm vir,t}<d_{\rm tn}<1.5R_{\rm vir,t}$) and even of the core spin vector ($d_{\rm tn}<0.8R_{\rm vir,t}$).
This can cause an unreal strong SOA.
To avoid such faulty measurement, we exclude close pairs with pairwise distances smaller than $0.8R_{\rm vir,t}$ in the sample.
The size of our sample is reduced to 84,014 neighbors in 70,766 target halos.

The next step is to quantify the strength of the SOA and its error, by adopting the method of \citet{Yang06}. 
For the analysis of three-dimensional alignment, we measure the directional cosine ($\cossl$) between the SOA's constituent vectors.
The frequency of interacting neighbors relative to the uniform distribution at a given spin--orbit angle is then defined as
\linenomath{
\begin{equation}
n(\cossl) \equiv \frac{N(\cossl)}{\langle N_{\rm rand}(\cossl) \rangle}\,, \label{eq:soasig} 
\end{equation}}
where $N(\cossl)$ is the number of neighbors at a given $\cossl$ and $\langle N_{\rm rand}(\cossl)\rangle$ is the mean value of $N_{\rm rand}(\cossl)$ obtained from 100 random isotropic samples with the same number of neighbors.
If the halo spin is preferentially aligned with the orbital angular momentum of the neighbor at a given $\cossl$, $n(\cossl)$ is greater than 1; $n(\cossl\approx1)>1$ ($\theta_{\rm SL}\approx0^{\rm \circ}$) is called {\it prograde alignment} and $n(\cossl\approx-1)>1$ ($\theta_{\rm SL}\approx180^{\rm \circ}$) is called {\it retrograde alignment}.
Random alignment is represented as $n(\cossl) = 1$ at all ranges of $\cossl$.

\section{Results} \label{sec:results}

\subsection{Spin--Orbit Angle Distribution}

\begin{figure}[ht!]
\centering
\vspace{-1.5ex}
\includegraphics[width=0.47\textwidth]{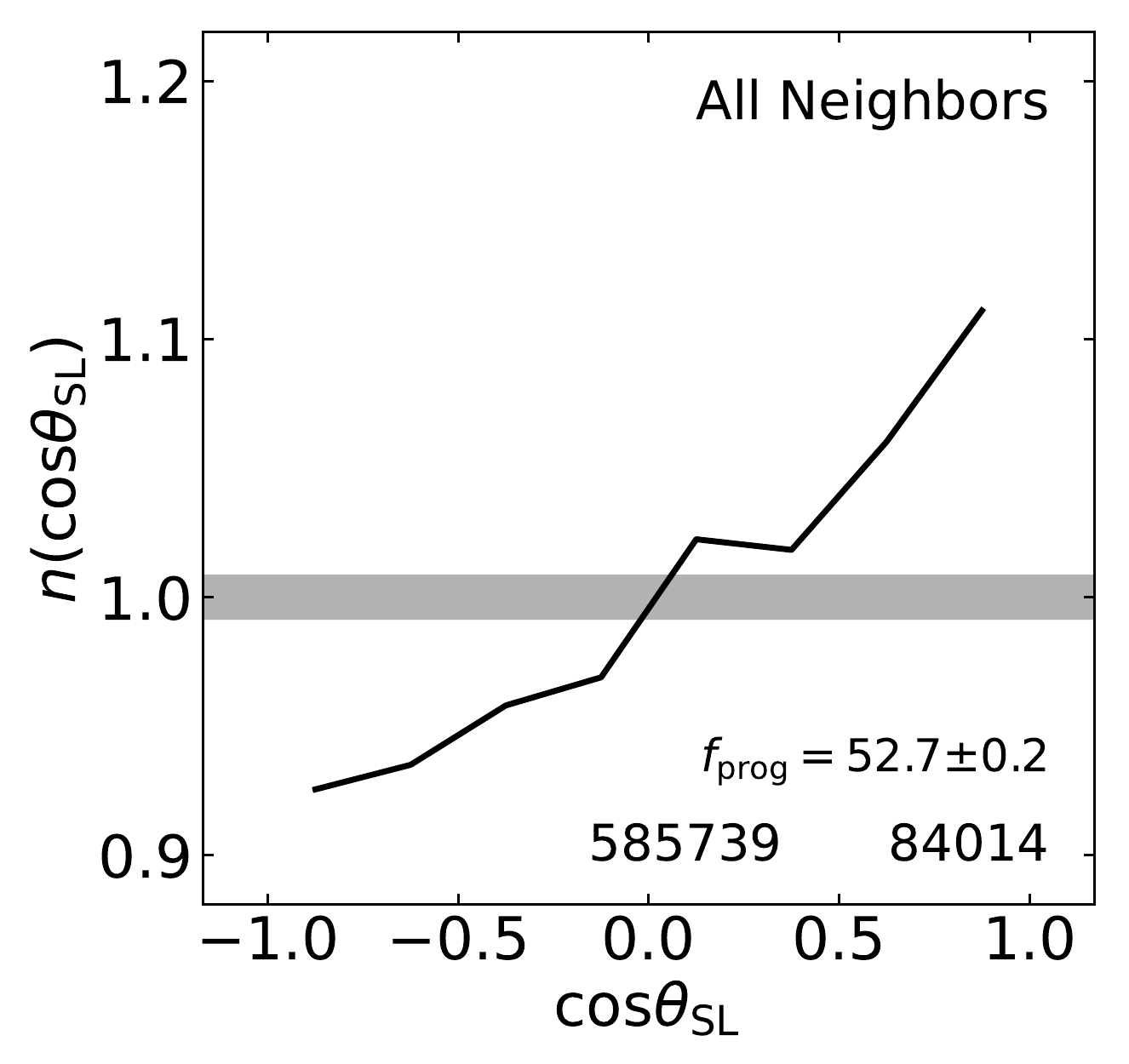}
\vspace{-5ex}
\caption{Spin--orbit angle distribution ($n(\cossl)$) for our whole target--neighbor pair sample at $z=0$. The horizontal shaded region represents the 1$\sigma$ uncertainty obtained from 100 randomly generated isotropic distributions with the same sample size. The value of $f_{\rm prog}$ is the prograde fraction (see Equation \ref{eq:fprog}). The numbers of all target halos (left) and their interacting neighbors (right) are given at the bottom. \label{fig:tsoasig}}
\vspace{-0.5ex}
\end{figure}

We examine the SOA for interacting neighbors of our target halos at $z=0$.
Figure \ref{fig:tsoasig} shows the spin--orbit angle distribution for our whole pair sample.
The alignment amplitude ($n(\cossl)$) increases with $\cossl$.
Compared to the random distribution, the number of highly prograding neighbors is 11\,\% higher and the number of highly retrograding neighbors is 8\,\% lower: $n(0.75<\cossl<1.0)=1.11\pm0.01$ and $n(-1.0<\cossl<-0.75)=0.92\pm0.01$.
The prograde alignment is thus noticeable while the retrograde alignment is insignificant.

To further demonstrate the existence of alignments, we measure the prograde fraction ($f_{\rm prog}$), which is defined as
\linenomath{
\begin{equation}
f_{\rm prog}\,(\%) \equiv \frac{N_{\rm n}(\cossl>0)}{N_{\rm n}(\rm total)} \times 100\,, \label{eq:fprog} 
\end{equation}}
where $N_{\rm n}$($\cossl>0$) is the number of interacting neighbors on the prograde orbit ($\cossl>0$) and $N_{\rm n}(\rm total)$ is the total number of interacting neighbors.
If the prograde fraction is greater than 50\,\% (a predicted value from the random distribution), prograde alignment exists.
For our pair sample, the prograde fraction is 52.7\,\%\,$\pm$\,0.2\,\%, indicating that significant prograde alignment is present with a significance level of $>13\,\sigma$.

\begin{figure*}[ht!]
\centering
\includegraphics[width=0.94\textwidth]{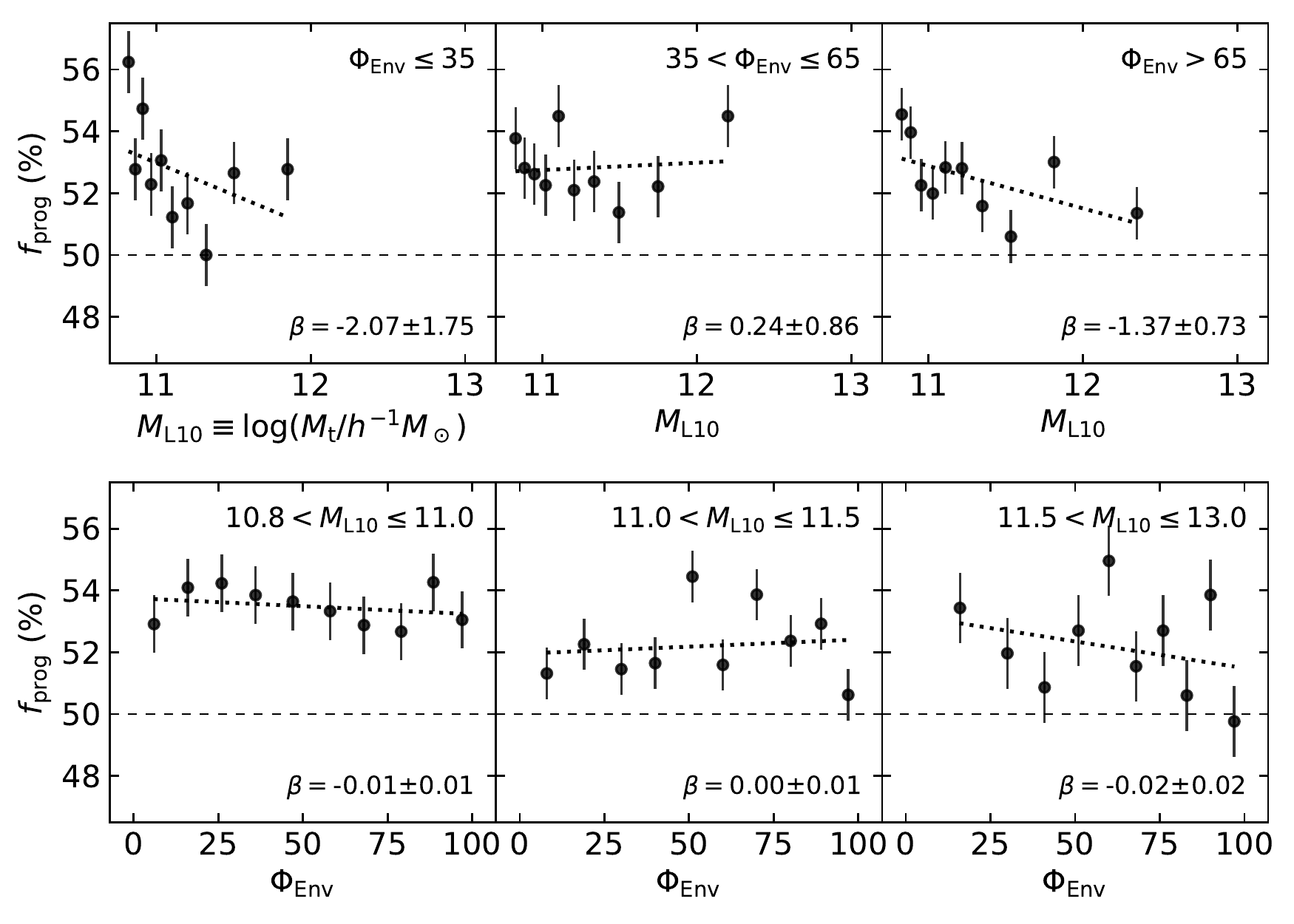}
\vspace{-2.5ex}
\caption{Prograde fraction ($f_{\rm prog}$) for all interacting neighbors. The top panels show the prograde fraction as a function of the target halo mass for three different environment subsamples. The bottom panels show the change of the fraction along with the environmental parameter for three different mass subsamples.
Each subsample has 10 bins with the same sample size.
In all panels, horizontal dashed lines at $f_{\rm prog}$ = 50\,\% indicate the value expected from the random distribution. 
The error bars along the $x$-axis are the standard error of the median (i.e., the ratio of the median absolute deviation to the square root of the sample size; $err(x)\equiv{\rm Med}(|x-{\rm Med}(x)|)/\sqrt{N_{\rm sample}}$) and they are all smaller than the symbol size. 
The error bars along the $y$-axis are the standard error from the binomial distribution ($err(y)\equiv\sqrt{f_{\rm prog}(1-f_{\rm prog})/N_{\rm sample}}$). 
Dotted lines show the linear fit to the subsamples using error-weighted orthogonal distance regression, and the slope ($\beta$) of the fitted line is given at the bottom of each panel. \label{fig:massenv}}
\vspace{1.5ex}
\end{figure*}

Such an alignment is consistent with previous findings from simulations \citep{Warn06,LHui17} and observations \citep{Herb08,LeeJ19a,LeeJ19b}.
\citet{LeeJ19b} found a coherence between the galaxy spin and neighbors' motion even up to a scale of several megaparsecs (800 kpc in \citealt{LeeJ19a}), beyond our distance criterion (on average $300\kpc$).
We can thus formulate two hypotheses: one is that SOA is caused by interactions with neighbors and the other is that SOA is brought about by the local LSS.
Our hypotheses are based on previous findings: (a) the flipping of the spin direction coincides with the interactions \citep{Bett12,Bett16}; (b) the halo spin is associated with the local LSS \citep{Arag07,Codi12,Temp13,Fore14,Zhan15,Wang18spin}; and (c) satellites are accreted along the local LSS \citep{Libe05,Kang15}.
In the following sections, we will investigate the dependence of SOA on various parameters and discuss the physical causes of SOA.

\subsection{Dependence of the Spin--Orbit Alignment \label{sec:soadep}}

\subsubsection{Dependence on the Halo Mass and Environmental Parameter \label{sec:massenv}}

Figure \ref{fig:massenv} shows the prograde fraction as a function of the halo mass (top panels) and the environmental parameter (bottom).
To analyze a pure mass and environmental dependence, we divide the sample into subsamples with respect to the halo mass and environment reducing the mass--environment relation.
Each subsample is broken down into 10 bins that have the same number of interacting neighbors rather than the same number of target halos because one target halo can have multiple neighbors.
We assess the mass and environmental dependence by fitting the prograde fractions of the subsamples using orthogonal distance regression.
The prograde fraction $f_{\rm prog}$ is described as simple linear functions, $f_{\rm prog}\propto\beta M_{\rm L10}$ and $\beta \Phi_{\rm Env}$, where $\beta$ is the slope of the trends.
The values of $\beta$ for the subsamples are given at the bottom of each panel.

For the given environmental parameters (top panels of Figure \ref{fig:massenv}), the prograde fraction tends to slightly decrease from $\sim$\,54\,\% to $\sim$\,51\,\% with increasing halo mass. 
The mass dependence of $f_{\rm prog}$ is statistically significant in low- and high-density environments (at the $1.2\sigma$ and $1.9\sigma$ levels, respectively), with little dependence for intermediate-density environments.
The decreasing trend is due to the fact that low-mass halos are so vulnerable to interactions with their neighbors that their spin vectors seem easily aligned with the neighbors' orbital angular momenta.
Such high $f_{\rm prog}$ for the low-mass halos is consistent with \citetalias{Moon21}.

For the given halo masses (bottom panels of Figure \ref{fig:massenv}), the prograde fraction is almost constant regardless of the large-scale density ($\lesssim$\,$1\sigma$).
This seems due to the fact that for the higher density, two competing factors arise: interactions with neighbors happen more frequently \citepalias[e.g.,][]{An19} but neighbors infall more isotropically \citep[e.g.,][]{Wang18sca}.
This weak dependence is consistent with that for high-mass galaxies in \citetalias[][]{Moon21}.
Although they found an environmental dependence on the SOA for less massive galaxies, their mass range is below ours since the mass resolution in our simulations is about 20 times lower than the TNG100's.

\subsubsection{Dependence on the Mass Ratio \label{sec:mratio}}

\begin{figure}[t!]
\centering
\includegraphics[width=0.47\textwidth]{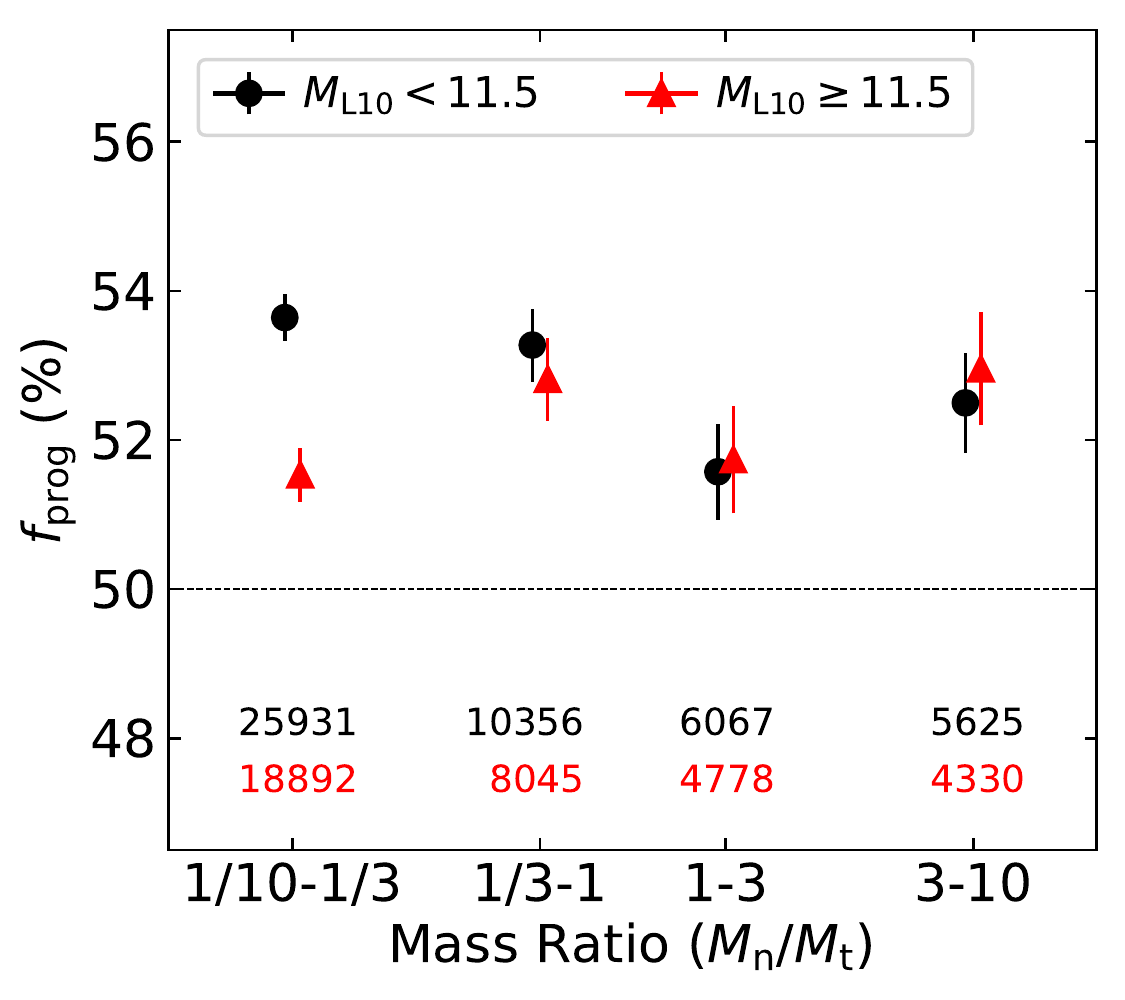}
\vspace{-5ex}
\caption{Same as Figure \ref{fig:massenv}, but with the prograde fraction as a function of mass ratio ($M_{\rm n}/M_{\rm t}$). To minimize the mass dependence of the prograde fraction, the sample is divided into two mass subsamples: lower-mass ($M_{\rm L10}<11.5$; black circles) and higher-mass ($M_{\rm L10}\geq11.5$; red triangles). The sample sizes for the four groups are given at the bottom of the panel. \label{fig:mratio}}
\end{figure}

Figure \ref{fig:mratio} shows the prograde fraction as a function of the mass ratio ($M_{\rm n}/M_{\rm t}$).
To minimize the mass dependence of $f_{\rm prog}$, we divide our sample into two subsamples: lower-mass ($M_{\rm L10}<11.5$) and higher-mass ($M_{\rm L10}\geq11.5$).
In the two mass ranges, the prograde fraction does not change significantly as the mass ratio varies.
Interestingly, for a mass ratio of 1/10--1/3 only, there is a clear difference in $f_{\rm prog}$ between the lower- and higher-mass halos.
The sample with this mass ratio range likely governs the mass dependence of $f_{\rm prog}$ in that $f_{\rm prog}$ for lower-mass halos is higher than that for higher-mass halos as shown in Figure \ref{fig:massenv}.

In Figure \ref{fig:mratio}, for the other mass ratio ranges, the values of $f_{\rm prog}$ for the low- and high-mass subsamples are the same within the errors.
This is due to (a) a dependence of the merger time scale on the mass ratio (for the mass ratio of 1/3--3) and (b) the incompleteness of identification of two close halos (for the mass ratio of 3--10).
First, the merging time for $1/3<M_{\rm n}/M_{\rm t}<3$ is shorter than that for $M_{\rm n}/M_{\rm t}<1/3$ and $M_{\rm n}/M_{\rm t}>3$ \citep{BT87,LC93,Boyl08,Jian08}.
A short interaction time is insufficient to bring about the mass dependence of $f_{\rm prog}$.
Next, a smaller target halo with $M_{\rm n}/M_{\rm t}>3$ seems to be embedded in its larger neighbor before the spin vector of the target is aligned with the neighbor's orbital angular momentum, and thus the prograde fraction appears not to depend on the mass.
This is due to the nature of the halo finding algorithm.
As the two halos get quite close to each other, the smaller halo is identified as a part of the larger neighbor rather than one separate halo \citep[e.g.,][]{Muld11}.
This hinders an examination of SOA caused by close neighbors with $M_{\rm n}/M_{\rm t}>3$ as the smaller halo spuriously disappears.

\subsubsection{Dependence on the Pairwise Distance \label{sec:dist}}

\begin{figure}[t!]
\centering
\includegraphics[width=0.47\textwidth]{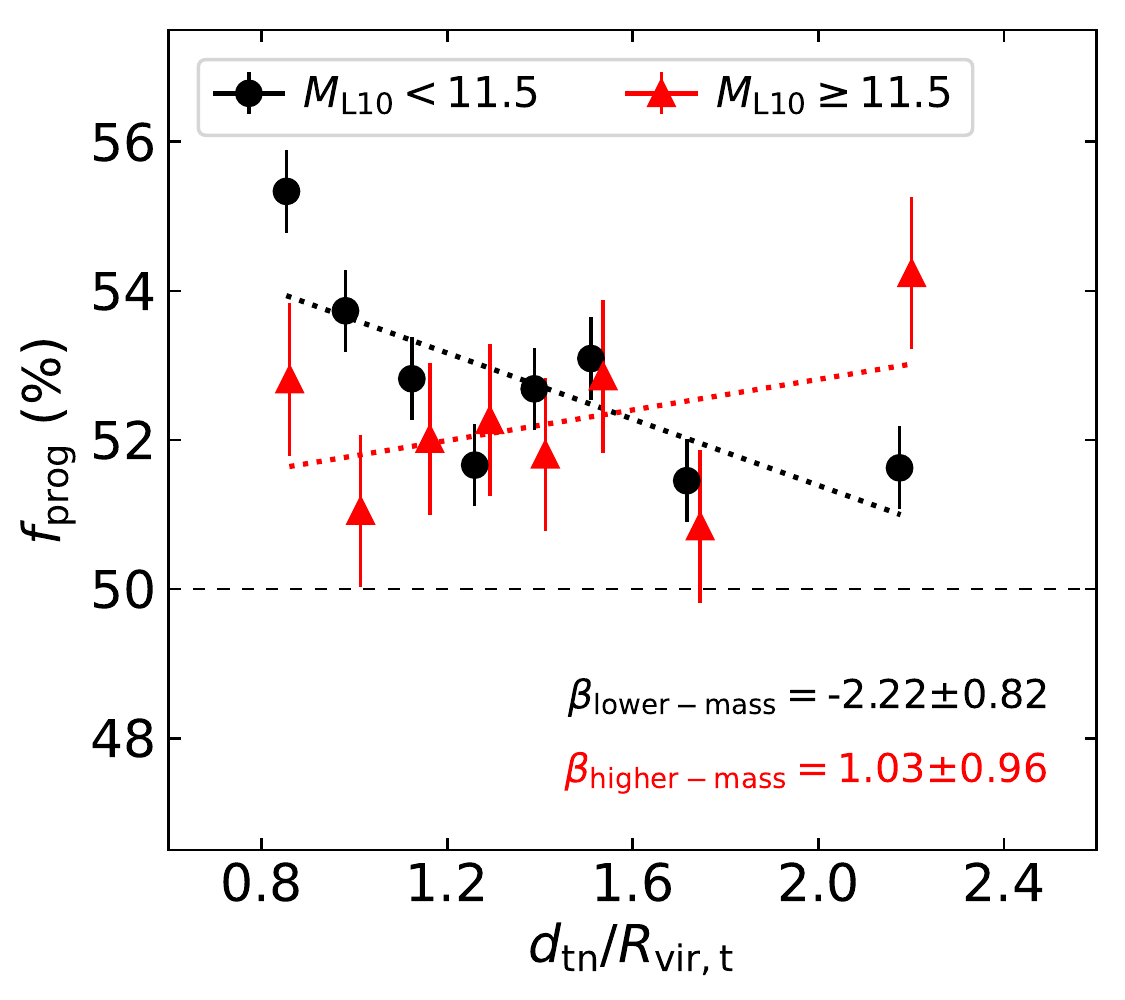}
\vspace{-5ex}
\caption{Same as Figure \ref{fig:massenv}, but with the prograde fraction as a function of normalized pairwise distance ($d_{\rm tn}/R_{\rm vir,t}$). For the same reason given in Figure \ref{fig:mratio}, the sample is divided into lower-mass ($M_{\rm L10}<11.5$; black circles) and higher-mass halos ($M_{\rm L10}\geq11.5$; red triangles). 
The subsamples have eight bins with the same sample size.
For each subsample, we fit the prograde fraction and give the slope ($\beta$) obtained from the linear fit at the bottom. \label{fig:pairdist}}
\end{figure}

Figure \ref{fig:pairdist} shows the prograde fraction as a function of the pairwise distance normalized to the virial radius of the target halo ($d_{\rm tn}/R_{\rm vir,t}$).
For the analysis, we divide our sample into lower-mass and higher-mass subsamples as in the previous section.
In addition, each subsample has eight bins with the same number of interacting neighbors.

We find a different trend between lower-mass ($M_{\rm L10}<11.5$) and higher-mass target halos ($M_{\rm L10}\geq11.5$).
For the lower-mass halos, as the pairwise distance increases, the prograde fraction decreases with a significance level of $2.7\sigma$.
This indicates that the halo spin is preferentially aligned with the orbital angular momentum of the neighboring halo in cases of close neighbors.
We note that such a preferential alignment simply implies a positive correlation on the ensemble average at $z=0$, not the evolution of the spin direction with the pairwise distance.
To confirm the time evolution of spins, we should use the merger tree data for each halo pair.

For the higher-mass halos ($M_{\rm L10}\geq11.5$), there is a marginally increasing trend of a $1.1\sigma$ significance level, and it is opposite to that for the lower-mass halos.
This is because, at our distance range, the spin orientation of the high-mass halos seems more difficult to change than that of the low-mass halos.
We speculate that, at $d_{\rm tn}<0.8R_{\rm vir,t}$, the dependence of $f_{\rm prog}$ on the pairwise distance would emerge if we could use an advanced method such as tracking particle IDs to separate a close pair into two halos.
On the other hand, the marginal increase at farther distance is likely associated with the LSS.
The halo mass $M_{\rm L10}=11.5$ is close to the spin-flip mass \citep[$M_{\rm flip}=10^{11.5-12.0}\hMsun$; e.g.,][]{Codi12,Gane18}.
For halos with a mass greater than the spin-flip mass, the spin directions have changed over cosmic time and become more linked to the cosmic flow through the filament.
This will have made the SOA stronger.
In Section \ref{sec:dndf} and \ref{sec:sfa}, we will further analyze the association of SOA with the LSS.

\begin{figure*}[ht!]
\centering
\includegraphics[width=0.935\textwidth]{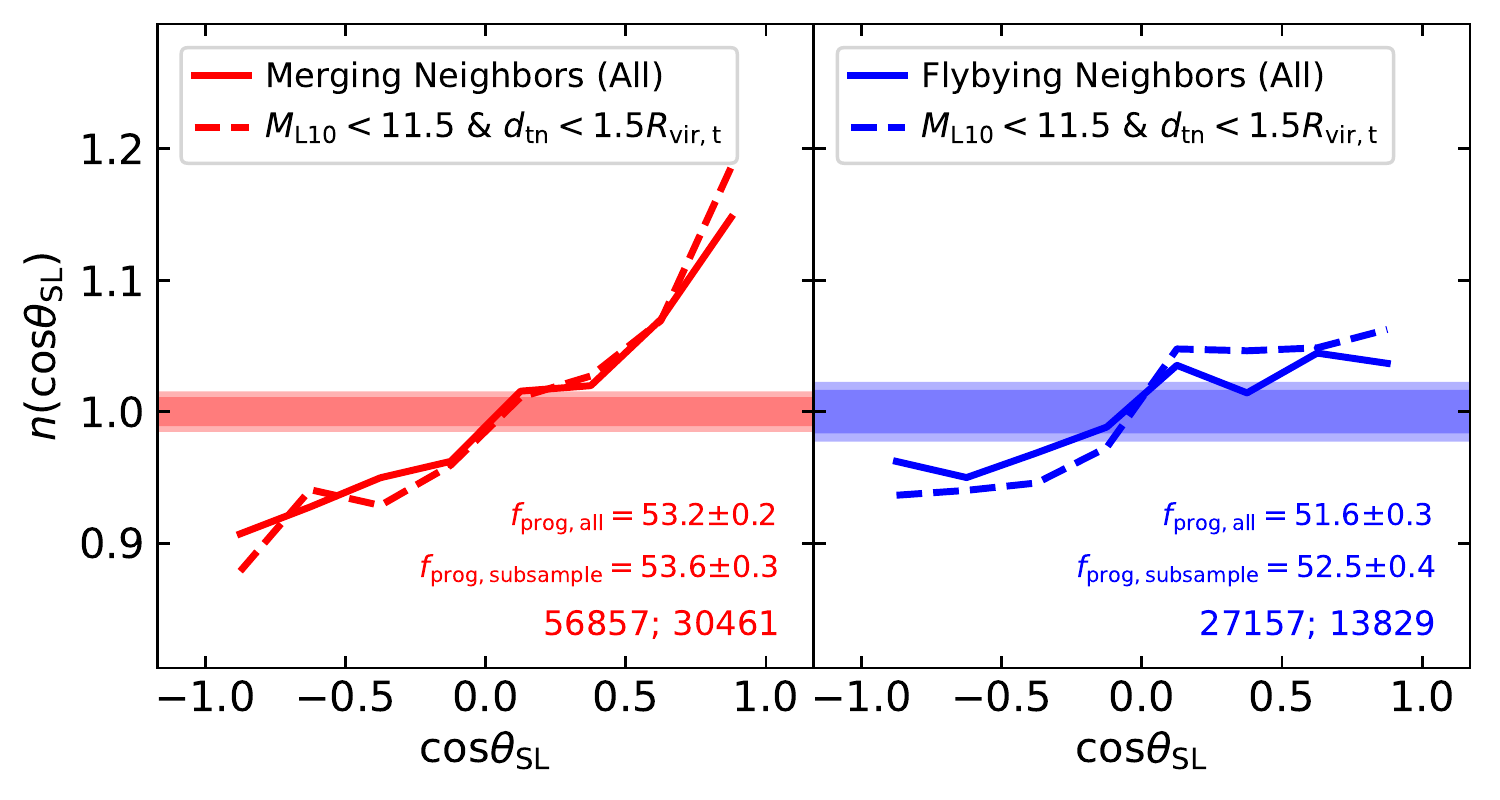}
\vspace{-2.5ex}
\caption{Same as Figure \ref{fig:tsoasig}, but for merging (left panel) and flybying (right) neighbors. The solid lines and dark shaded regions are for our whole sample. The dashed lines and light shaded regions are for a subsample with $M_{\rm L10}<11.5$ and $d_{\rm tn}<1.5R_{\rm vir,t}$, which shows strong SOA. 
The prograde fractions and the numbers of targets from the whole sample and subsample are given at the bottom of the panels. \label{fig:soamgfb}}
\vspace{1.4ex}
\end{figure*}

\begin{figure*}[ht!]
\centering
\includegraphics[width=0.935\textwidth]{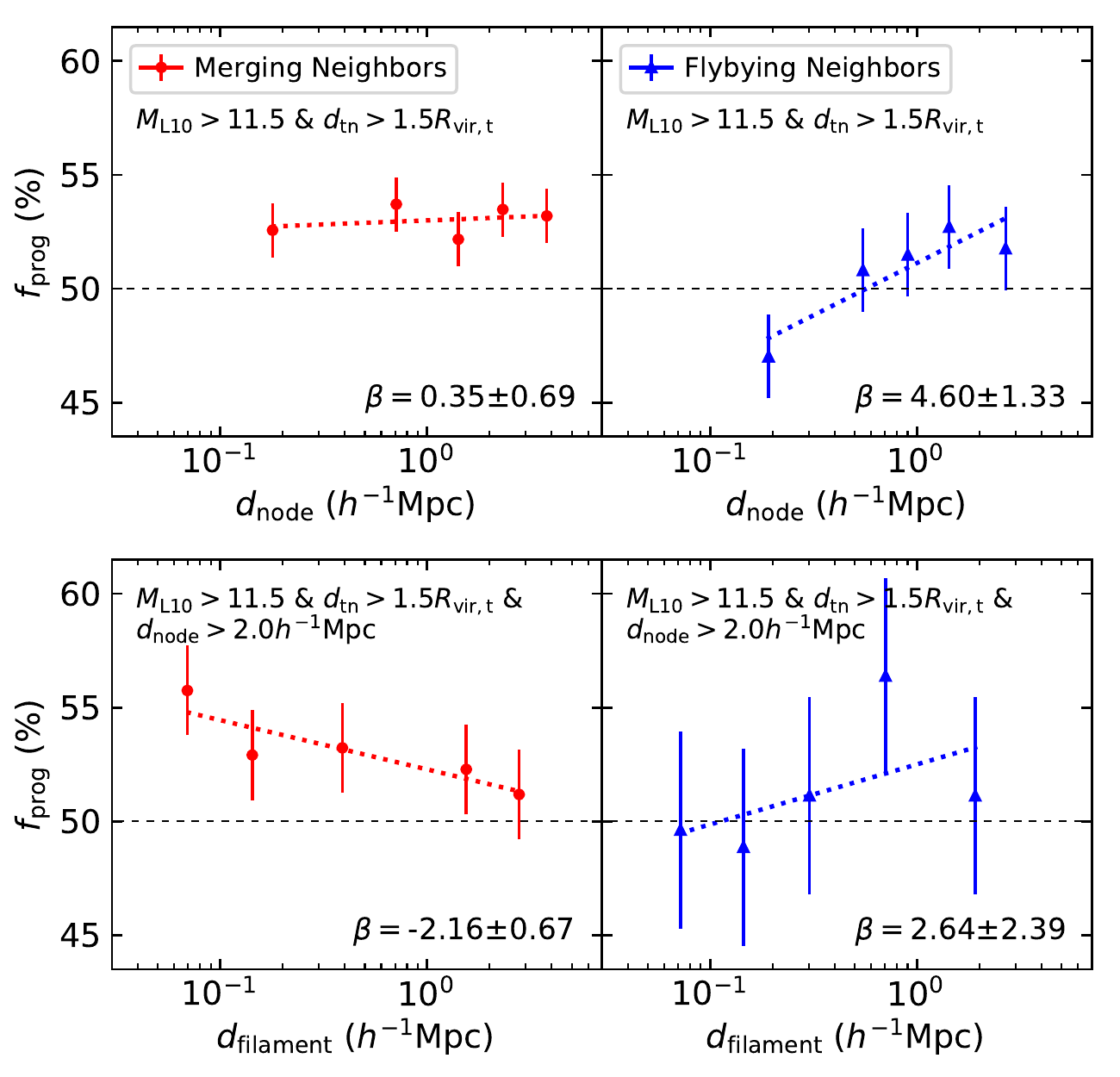}
\vspace{-2.8ex}
\caption{Same as Figure \ref{fig:massenv}, but with the prograde fraction as a function of the distances to the nearest node ($d_{\rm node}$; top panels) and filament ($d_{\rm filament}$; bottom panels). To minimize the neighbor-interaction effect, our sample is restricted to $M_{\rm L10}\geq11.5$ and $d_{\rm tn}>1.5R_{\rm vir,t}$.
In the bottom panels, to examine the refined dependence on $d_{\rm filament}$, the sample is more tightly restricted to $d_{\rm node}>2.0h^{-1}{\rm Mpc}$ than that in the top panels.
The samples have five bins with the same sample size.
The slope ($\beta$) obtained from the linear fit of each subsample is given at the bottom of each panel. \label{fig:dndf}}
\vspace{1.4ex}
\end{figure*}

\subsubsection{Dependence on the Interaction Type \label{sec:mgfb}}

In Figure \ref{fig:soamgfb}, we compare the spin--orbit angle distributions for merging neighbors (left panel) and flybying neighbors (right).
The merger sample shows a stronger prograde alignment than the flyby sample.
As the directional cosine of the spin--orbit angle increases, the difference in $n(\cossl)$ between the merger and flyby samples turns from negative ($0.91-0.96=-0.05$) to positive ($1.15-1.03=0.12$).
In other words, merging neighbors are more preferentially on the prograde orbit than flybying neighbors while the flybying neighbors' orbits are closer to being random.
This trend holds true even for the subsample with $M_{\rm L10}<11.5$ and $d_{\rm tn}<1.5R_{\rm vir,t}$ that shows a strong SOA in Sections \ref{sec:massenv} and \ref{sec:dist}.
In the whole sample (subsample), the prograde fraction for merging neighbors is 1.6\,\% (1.1\,\%) higher than that for flybying neighbors with a significance level of $5\sigma$ ($3\sigma$).
We attribute the stronger prograde alignment for merging neighbors to their duration of interaction being longer than that of flybying neighbors.
In other words, flybying neighbors mainly have a shorter interaction duration to exchange the tidal torque, due to their higher relative velocities compared to those of merging neighbors \citepalias[e.g.,][]{An19}.

On the other hand, the small difference in the spin--orbit angle distribution between the whole sample and the subsample indicates that the stronger SOA for merging neighbors compared to the SOA for flybying ones can be detected even at a farther distance.
This is a natural feature according to two facts: (a) the spin direction is induced by the tidal effect of the local cosmic flow \citep[e.g.,][]{Dubi92,Fore14} and (b) distant merging neighbors better reflect the local flow than close ones do \citep[e.g.,][]{Welk18}.
A merger thus forms SOA by coming along the flow even from the beginning of the interaction.
By contrast, a flybying neighbor would move irrespective of the flow, and such movement seems to imprint a nearly flat feature in the flyby's spin--orbit angle distribution.
Hence, the SOA for flybying neighbors is inherently weak.

\begin{figure*}[ht!]
\centering
\includegraphics[width=0.93\textwidth]{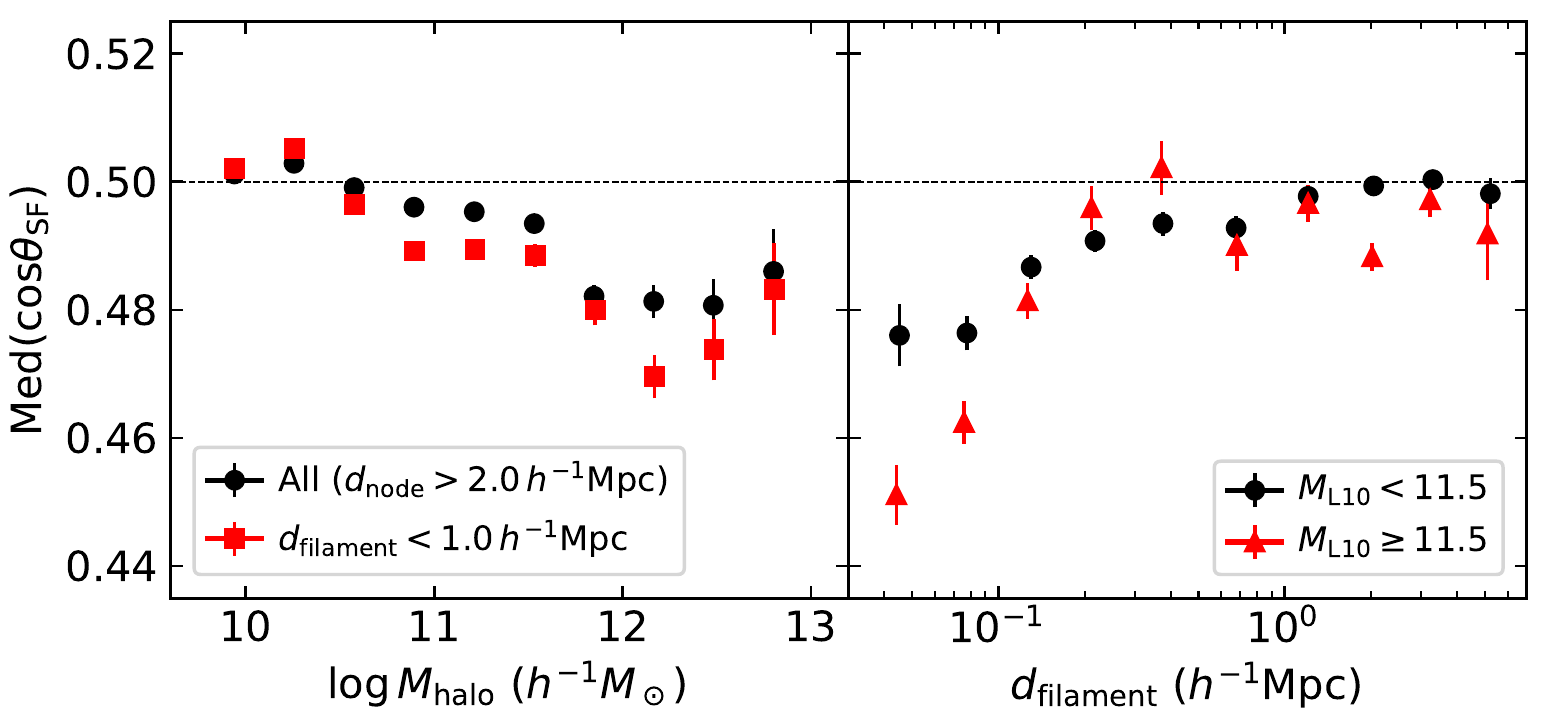}
\vspace{-1.5ex}
\caption{Degree of spin--filament alignment as a function of the halo mass (left panel) and the distance to the nearest filament (right). To better focus on the correlation between the halo spin and filament, the sample is restricted to $d_{\rm node}>2.0\Mpc$. 
In the left panel, black circles and red squares represent the entire sample and the filament sample ($d_{\rm filament}<1.0\Mpc$), respectively. 
The subsamples have 10 bins with an equal width along the $x$-axis.
In the right panel, the black circles and red triangles are the lower- and higher-mass halos, respectively. 
The subsamples have 10 bins with an equal log-scale width.
In both panels, the error bars are the standard error of the median, and some symbols with a large sample size have errors smaller than the symbol size. 
\label{fig:sfalign}}
\vspace{1.4ex}
\end{figure*}

\subsubsection{Dependence on the Distance to the Nearest Node and Filament \label{sec:dndf}}

In this section and the next, we focus on the contribution of the LSS to the SOA.
For a pure analysis, we minimize the neighbor-interaction effect on the SOA by restricting our sample to the subsample with $M_{\rm L10}\geq11.5$ and $d_{\rm tn}>1.5R_{\rm vir,t}$, which has the weakest prograde alignment.
We also compare the LSS effects between merging and flybying neighbors since they may have different origins as mentioned in Section \ref{sec:mgfb}.
The subsample size is reduced once again for an examination of dependence on the distance from the nearest filament because we exclude the halos residing around the node within $2.0\Mpc$ as used in previous studies on filaments \citep[e.g.,][]{Kral18}.

Figure \ref{fig:dndf} shows the prograde fraction as a function of the distances from the nearest node ($d_{\rm node}$; top panels) and the nearest filament ($d_{\rm filament}$; bottom).
In the top left panel, the prograde fraction for merging neighbors does not depend on $d_{\rm node}$ ($<1\sigma$) and remains constant at $\sim$\,53\,\%.
Such a constant $f_{\rm prog}$ seems to suggest that merging neighbors come along the same filament and interact with their target halos.
This is in line with the observation that the major axes of Virgo cluster galaxies are aligned with the axes of filaments connected to the cluster \citep{KimS18}.
In the top right panel, the prograde fraction for flybying neighbors increases with $d_{\rm node}$ ($3.5\sigma$).
In other words, flybying neighbors closer to the node have a more random incidence angle when entering their target halo.
This suggests that flybying neighbors come from different filaments.
The node is an environment involved with a number of filaments.
If one neighbor moves along a filament different from its own and encounters its target halo, its relative velocity is so high that the neighbor will be a flybying neighbor.
Since filaments are manifold around the node, the SOA would disappear.

In the bottom left panel of Figure \ref{fig:dndf}, the prograde fraction for merging neighbors decreases with distance from the nearest filament.
The trend has a significance level of $3.2\sigma$.
The higher prograde fraction for halos closer to the filament has two possibilities: (a) if the spin vector of the halo is parallel to the filamentary spine, its neighbor moves perpendicularly to the spine, or (b) if the spin vector is perpendicular to the filament, the neighbor moves along the spine.
The halos of our subsample are massive ($>10^{11.5}\hMsun$) and thus their spin orientations are largely perpendicular to the filament \citep[e.g.,][]{Codi12}.
Hence, the second possibility seems more plausible.
In the bottom right panel, for flybying neighbors, $f_{\rm prog}$ decreases with decreasing $d_{\rm filament}$.
This is consistent with the trend found between $f_{\rm prog}$ and $d_{\rm node}$ although the significance level is $1.1\sigma$.
The low $f_{\rm prog}$ at small $d_{\rm filament}$ is due to the center of the filament being an intricate intersection of cosmic flow like the node environment.

The correlation of $f_{\rm prog}$ with $d_{\rm node}$ and $d_{\rm filament}$ is a combination between the neighbor-interaction effect and the pure node and filament effect.
Due to the high number density of halos around the node and filament, interactions with multiple neighbors are frequent, burying the node and filament effect on the SOA under the neighbor-interaction effect.
Also our criterion of $d_{\rm tn}>1.5R_{\rm vir,t}$ excludes target--neighbor pairs residing close to the node and filament, which prohibits examining the dependence of the SOA on their environment.
To disentangle the two effects, we will scrutinize the SOA with respect to the cosmic web structure.
In the following section, we will further investigate the SOA's correlation with the angular alignment between the spin direction and the filament axis given that the filamentary structure is anisotropic unlike the spherically symmetric node.

\subsubsection{Dependence on the Degree of Spin--Filament Alignment \label{sec:sfa}}

Figure \ref{fig:sfalign} shows the median of the directional cosine (i.e., the spin--filament angle $\cos\theta_{\rm SF}$) between the spin vector of a halo ($\vect{S}$) and the axis of its nearest filament ($\vect{F}$) as a function of the halo mass (left panel) and the distance from the nearest filament (right).
To see the pure correlation of the halo spin with the filament, we exclude the node sample with $d_{\rm node}<2.0\Mpc$.

In the left panel, the median value of $\cos\theta_{\rm SF}$ decreases with the halo mass.
The trend is consistent with previous findings in both simulations \citep[e.g.,][]{Arag07,Codi12,Gane18} and observations \citep[e.g.,][]{Welk20}: the lower-mass (higher-mass) halo prefers to have a spin parallel (perpendicular) to the nearest filamentary spine.
The mass for which the median value crosses 0.5 is approximately $10^{11}\hMsun$.
This is smaller than the well-known spin-flip mass ($10^{11.5-12.0}\hMsun$; e.g., \citealt{Codi12}) and is attributed to the use of the inner halo spin.
\citet{Gane18} found that the spin-flip mass for the inner halo spin is lower than that for the whole halo spin.
We have verified that the spin-flip occurs at $\sim$\,$10^{11.5}\hMsun$ when considering the whole halo spin.
The spin-flip phenomenon holds true for halos close to the filament ($d_{\rm filament}<1.0\Mpc$).
Their median value of $\cos\theta_{\rm SF}$ is slightly smaller than that for the whole sample.
This indicates that the spin-flip happens more frequently as the halos are closer to the filament.

In the right panel of Figure \ref{fig:sfalign}, the degree of spin--filament alignment depends on the distance from the nearest filament as well.
The median value of $\cos\theta_{\rm SF}$ increases with $d_{\rm filament}$ regardless of the halo mass.
This indicates that, as $d_{\rm filament}$ increases, the averaged halo spin vector turns from being perpendicular to the filament into being parallel to the filament.
In view of the mass dependence of the spin--filament alignment, the spin vectors of lower-mass halos are thought to be parallel to the filament.
However, lower-mass halos close to the filament ($d_{\rm filament}\lesssim1.0\Mpc$) have experienced the spin-flip, despite a median value of $\cos\theta_{\rm SF}$ that is higher than that for higher-mass halos.
The difference in the fraction of spin-flipped halos between two mass ranges appears to generate the mass dependence of the spin--filament alignment.
Hence, the halo spin vector is largely perpendicular to the filament when the halo resides close to the filament.

\begin{figure*}[ht!]
\centering
\includegraphics[width=0.94\textwidth]{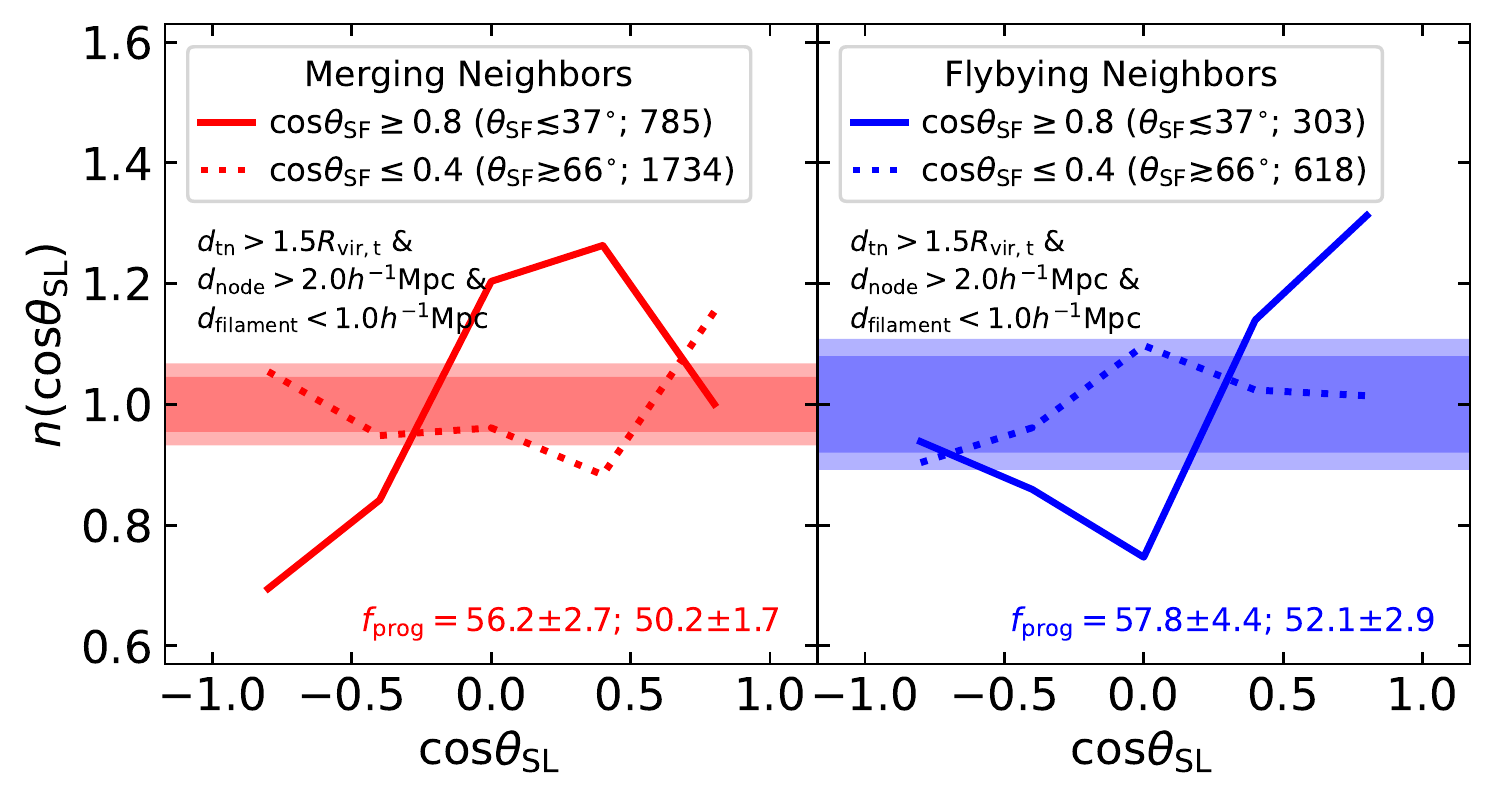}
\vspace{-2ex}
\caption{Same as Figure \ref{fig:tsoasig}, but for the target halo sample with distant neighbors ($d_{\rm tn}>1.5R_{\rm vir,t}$) and in the filament ($d_{\rm node}>2.0\Mpc$ and $d_{\rm filament}<1.0\Mpc$). Based on the spin--filament angle ($\cos\theta_{\rm SF}$), the sample is divided into two subsamples: `parallel' ($\cos\theta_{\rm SF}\geq0.8$ or $\theta_{\rm SF}\lesssim37^{\circ}$; solid lines) and `perpendicular' ($\cos\theta_{\rm SF}<0.4$ or $\theta_{\rm SF}\gtrsim66^{\circ}$; dotted lines). 
The sizes of the subsamples are given in the legend box and the prograde fractions are given at the bottom of the panels. \label{fig:soasf}}
\vspace{1.5ex}
\end{figure*}

Figure \ref{fig:soasf} shows the spin--orbit angle distribution depending on the spin--filament angle. 
We use the filament sample ($d_{\rm node}>2.0\Mpc$ and $d_{\rm filament}<1.0\Mpc$) with far neighbors ($d_{\rm tn}>1.5R_{\rm vir,t}$).
The sample is divided into two subsamples: {\it parallel} ($\cos\theta_{\rm SF}\geq0.8$, i.e., $\theta_{\rm SF}\lesssim37^{\circ}$) and {\it perpendicular} ($\cos\theta_{\rm SF}\leq0.4$, i.e., $\theta_{\rm SF}\gtrsim66^{\circ}$) with respect to the filament axis used in \citet{Welk18}.

For merging neighbors (left panel), the parallel subsample has a maximum amplitude at $\cossl\sim0.4$.
The excess at $0.2<\cossl<0.6$ is significant: $n(\cossl)=1.26\pm0.07$.
At $\cossl>0.4$, the amplitude dramatically decreases with $\cossl$.
By contrast, the perpendicular subsample shows a flat spin--orbit angle distribution with a slight increase at both ends.
The prograde fraction for the perpendicular subsample is lower than that for the parallel subsample.
This indicates that the halos of the parallel subsample interact largely with neighbors on a nearly perpendicular but prograde orbit.
Hence, the neighbors coming along the filament lead to a high prograde fraction for halos close to the filament.
The difference between the two merger subsamples appears to represent different states before and after the spin-flip.
Therefore, if the spin vector of a halo is parallel to the filament, the halo confronts the spin-flip soon by its neighbor with an orbital angular momentum perpendicular to the spin.
After the spin-flip, the SOA at $\cossl>0.5$ develops as the frequency of the nearly perpendicular interactions decreases.

For flybying neighbors (right panel of Figure \ref{fig:soasf}), the parallel subsample has a different spin--orbit angle distribution from that for merging neighbors.
The distribution has a minimum value at $\cossl=0.0$ and dramatically increases.
Such a prograde alignment diminishes for the perpendicular subsample.
Their high relative velocities (\citealt{Gned03}; \citetalias{An19}) imply that flybying neighbors move along the fastest-collapsing direction.
When the spin vector of a halo is parallel to the filament, it should be aligned with the orbital angular momentum of the flybying neighbor.
The filament axis is however the slowest-collapsing direction under the definition of the cosmic web \citep{Caut14}.
A halo with a spin perpendicular to the filament encounters neighbors coming both from the outskirts of the filament and along the filament, and thus the SOA is broken.

\begin{figure*}[ht!]
\centering
\includegraphics[width=0.93\textwidth]{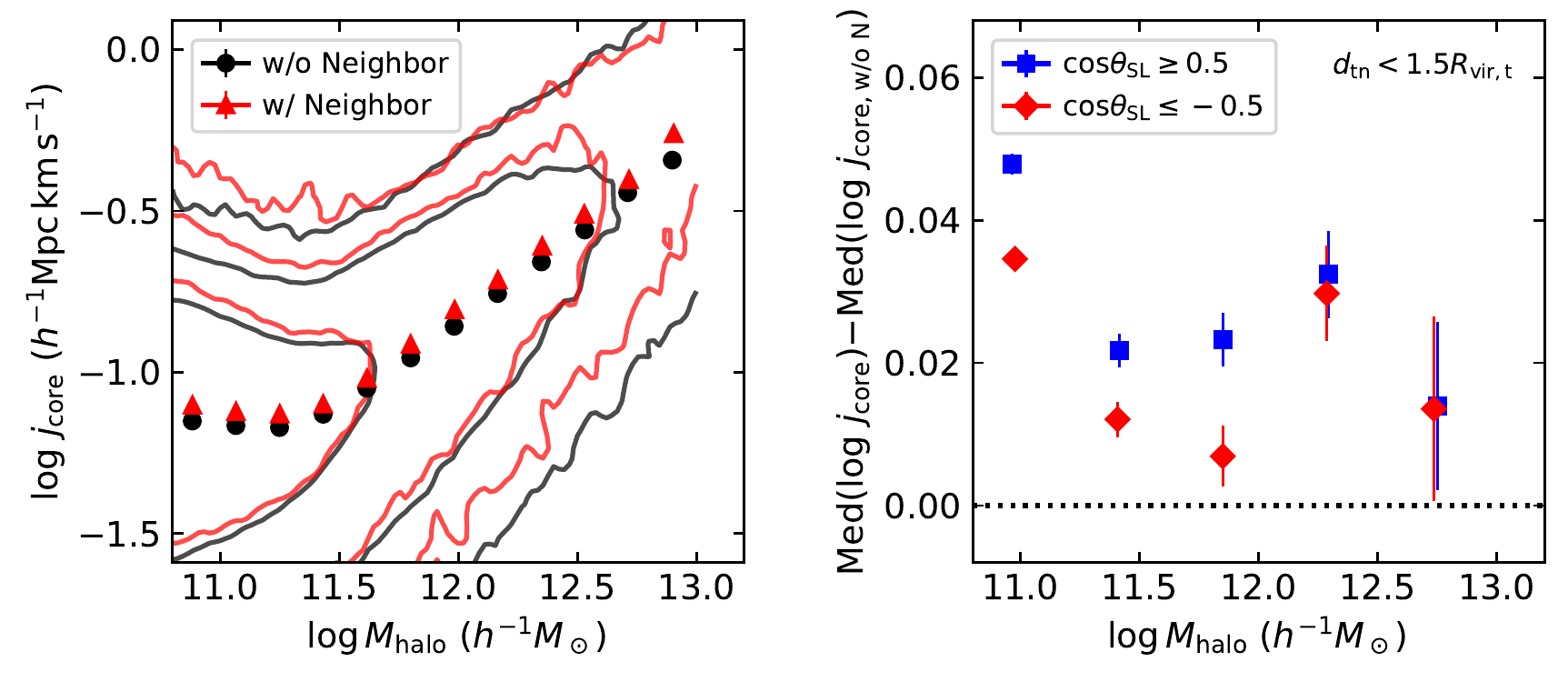}
\vspace{-2ex}
\caption{Left: Specific angular momentum ($j_{\rm core}\equiv |\vect{S}|/M_{\rm halo}$) of the core region of halos as a function of the halo mass. We compare the median values of $j_{\rm core}$ between two subsamples: halos without neighbors (black circles) and halos with neighbors (red triangles). 
The subsamples have 12 bins with an equal width along the $x$-axis. The error bars are the standard error of the median and they are all smaller than the symbol size.
Contours denote $1\sigma$, $2\sigma$, and $3\sigma$ of the subsamples' distributions.
Right: Difference in the median value of $j_{\rm core}$ between halos with neighbors (symbols) and those without neighbors (dotted horizontal line). The halos with neighbors are divided into two subsamples: highly prograde ($\cossl\geq0.5$; blue squares) and highly retrograde ($\cossl\leq-0.5$; red diamonds).
The subsamples have five bins with an equal width along the $x$-axis. The error bars are the standard error of the median, and those along the $x$-axis are all smaller than the symbol size.
\label{fig:specam}}
\vspace{1.5ex}
\end{figure*}

\section{Discussion}

We have shown that the spin vector of a halo is strongly aligned with the orbital angular momentum vector of its neighbor (see Figure \ref{fig:tsoasig}).
We have found three interesting facts about SOA: (1) halos with close neighbors show stronger SOA than do halos with far neighbors (see Figure \ref{fig:pairdist}); (2) merging pairs show stronger SOA than do flybying pairs (see Figure \ref{fig:soamgfb}); and (3) halos close to the filament show stronger SOA than do halos far from the filament (see Figure \ref{fig:dndf}).
The first two suggest that the formation of SOA is led by the interaction, whereas the last one implies that it is led by the LSS.
We discuss the link between SOA and the two physical causes: interaction with neighbors (Section \ref{sec:discussint}) and the LSS (Section \ref{sec:discusslss}).

\subsection{Linking SOA to Interaction with Neighbors \label{sec:discussint}}

SOA is stronger for halos with closer neighbors than for those with farther neighbors (see Figure \ref{fig:pairdist}).
We attribute this to the tidal effect of interactions with neighbors.
With a closer neighbor, the neighbor's orbital angular momentum could modify the target's spin vector to be more aligned with itself.
This is because the magnitude of the neighbor's orbital angular momentum is on average greater than the internal angular momentum of the target \citep{Hetz06}.
The neighbor-interaction effect on the SOA holds true for the SOA of galaxies \citepalias{Moon21}.
In the same vein, \citet{Cerv10} found that the spin directions of two galaxies in a pair are more parallel for the smaller pairwise distance.
\citet{LeeJ20} showed that paired galaxies with stronger SOA physically resemble each other more (in terms of color) than the other paired galaxies with weaker SOA.
These results imply that interacting neighbors affect a number of physical properties of their target halos.
We note that the stronger SOA for closer neighbors is detectable for low-mass target halos because they are more sensitive to the interaction than high-mass halos.
This entails a mass dependence on the SOA (see Figure \ref{fig:massenv}).
Despite the weak SOA for high-mass halos, the interaction effect on the SOA would be more prominent if we could use neighbors with distances shorter than our criterion of $0.8R_{\rm vir,t}$.

Merging neighbors make the SOA stronger than flybying neighbors do.
Although both types of interaction tidally affect the target halos, there is a difference in the interaction duration between mergers and flybys.
The duration of mergers is longer than that of flybys.
A merger exerts a strong tide during an interaction and converts the orbital angular momentum of a neighbor into the spin of its target halo \citep{Hetz06}, and then the two vectors become aligned \citep{Fern17}.
We thus suggest that the SOA grows mainly by prolonged interactions with merging neighbors.
By contrast, a flyby has a shorter duration of interaction than a merger due to a relatively rapid velocity (\citealt{Gned03}; \citetalias{An19}). 
The instantaneous interaction causes only a limited conversion from the external angular momentum into the internal.

In Figure \ref{fig:specam}, we further examine the effect of the neighbor-interaction on the halo spin.
We measure the specific angular momentum ($j_{\rm core}\equiv|\vect{S}|/M_{\rm halo}$) of the core region of a halo, where $\vect{S}$ is the internal angular momentum of the halo core as used in the analysis of this study.
The left panel shows the specific angular momentum as a function of the halo mass.
Compared to halos without neighbors, those with neighbors have, on average, a slightly higher $j_{\rm core}$, and it is consistent with the previous findings \citep[e.g.,][]{LL13,John19}.
Although the increase in $j_{\rm core}$ is statistically significant, the difference between the two subsamples is quite small.
This is because there are multiple drivers of the rise in $j_{\rm core}$, such as the halo mass, pairwise distance, interaction type, and degree of SOA.

In the right panel of Figure \ref{fig:specam}, we investigate how much the specific angular momentum increases depending on the degree of the SOA, which is one of the drivers of the increase in $j_{\rm core}$.
Compared to the median value of $j_{\rm core}$ for halos without neighbors, the median value of $j_{\rm core}$ is higher for halos with neighbors, regardless of their orbital configuration.
This seems due to the fact that the presence of interacting neighbors is associated with the growth of $j_{\rm core}$ via matter accretion along the local cosmic flow \citep{John19}.
The result rather differs from our initial expectation that prograde interaction raises $j_{\rm core}$ but retrograde interaction lowers it.
Nonetheless, highly prograding neighbors ($\cossl\geq0.5$) enhance $j_{\rm core}$ more than highly retrograding neighbors ($\cossl\leq-0.5$).
Such an increasing trend gets stronger for lower-mass halos because they are so susceptible that their spin vectors can easily change.
The fast rotation induced by prograde interaction may hold true not only for dark matter halos but for galaxies \citep{Lago18}.
Specifically for a disk galaxy, continuous prograde accretion is required to maintain the angular momentum of its disk \citep{Sale12,PR20}.

Interacting neighbors tidally affect both the direction and magnitude of the halo spin.
Whereas the spin direction becomes aligned with the specific direction during the interaction, the spin magnitude can either increase or decrease depending on the orbital configuration.
This makes it difficult to detect the neighbor-interaction effect on the spin magnitude.
This holds true even in view of the spin parameter ($\lambda$; \citealt{Peeb69,Bull01}).
\citet{Hetz06} found that $\lambda$ instantaneously increases during the merger process but then decreases to even lower than its value before the merger.
Merger and flyby interactions ultimately lead to slow-rotating halos \citep[e.g.,][]{Cape11} by reducing the internal energy of the halo \citep{Hetz06}.
Quantifying the interaction effect on the SOA and the halo spin will require individually tracing the direction and magnitude of the halo spin during the interactions.

\begin{figure*}[ht!]
\centering
\includegraphics[width=\textwidth]{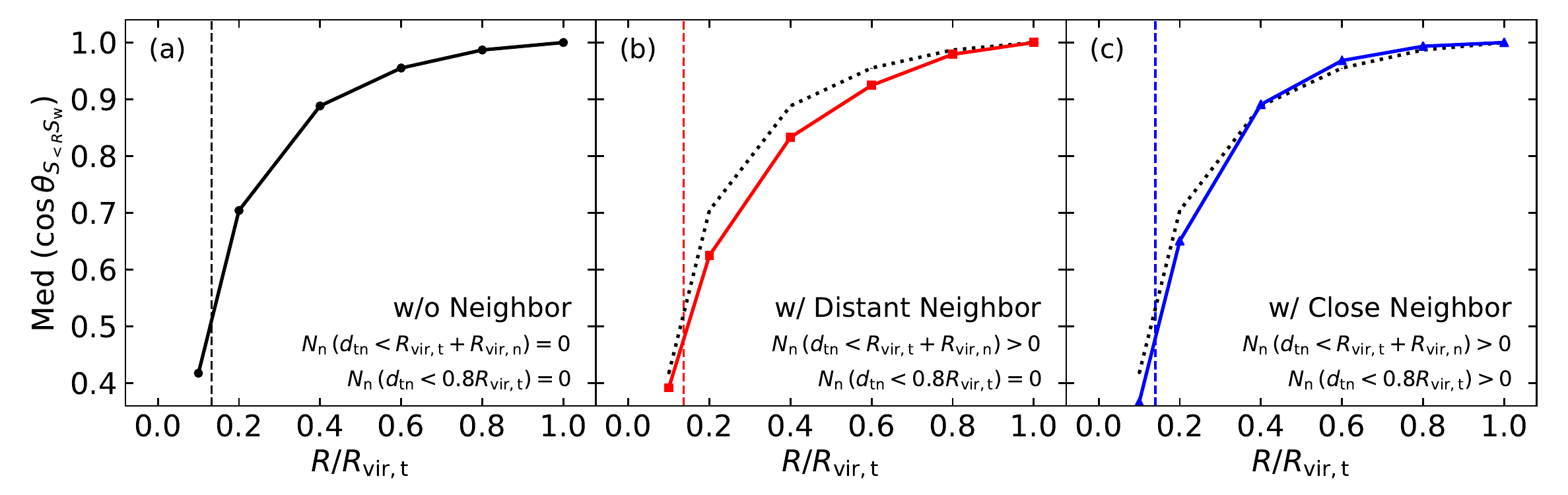}
\vspace{-5ex}
\caption{Offset angles between the spin direction of each target halo measured within a specific radius ($\vect{S_{<R}}$) and that measured within the virial radius ($\vect{S_{\rm w}}$). The target halo sample is divided into three subsamples: (a) target halos without neighbors that have a distance smaller than $R_{\rm vir,t}+R_{\rm vir,n}$, (b) those with distant neighbors that meet the distance criterion imposed in our target--neighbor pair sample ($0.8R_{\rm vir,t}<d_{\rm tn}<R_{\rm vir,t}+R_{\rm vir,n}$), and (c) those with close neighbors that have a distance smaller than our minimum distance criterion ($0.8R_{\rm vir,t}$). The halo spin directions are measured within 0.1, 0.2, 0.4, 0.6, 0.8, and 1.0 times the virial radius. Errors are the standard error of the median value but are smaller than the symbol size. The dotted lines in (b) and (c) are for the target halos without neighbors shown in (a). Vertical dashed lines represent the median value of the core radii of the subsamples, and the values are approximately $0.13R_{\rm vir}$.
\label{fig:SpinRadius}}
\vspace{1.5ex}
\end{figure*}

\subsection{Linking SOA to the LSS \label{sec:discusslss}}

This paper is the first of its kind to demonstrate the link of SOA to the LSS.
We propose that the LSS is one of the physical causes of SOA based on the tidal torque theory \citep{Peeb69}.
As the tidal torque by the primordial density fluctuation regulates the halo spin direction to be perpendicular to the cosmic flow \citep[e.g.,][]{Wang11}, SOA can emerge naturally.
Our results show that the halo spin is well aligned even with a distant neighbor's orbital angular momentum, which represents the local cosmic flow along the filament for merging neighbors and toward the filament for flybying neighbors.
In other words, the regulation of the spin direction by the LSS entails SOA.

We examined the correlation of SOA with the degree of spin--filament alignment (see Figure \ref{fig:soasf}).
Most interestingly, we discovered for the first time that halos with a spin parallel to the filament experience most frequently prograde-polar merging interactions with $\cossl\simeq0.4$ (i.e., $\theta_{\rm SL}\sim70^{\circ}$).
This is a piece of evidence supporting the notion that merging neighbors move along the filament, as shown in previous studies \citep{Libe11,Shi15,Muss18}.
This is logically consistent with \citet{TT15} and \citet{Mesa18}, who observed that the orientation of galaxy pairs is aligned with the spine of the nearest filament.
The distribution of farther satellites is also better aligned with the filament than with the major axis of their centrals \citep{Dong14,Welk17,Welk18}.
We suggest that at the beginning of the interaction, the orbital angular momentum of a neighbor representing the cosmic flow is aligned with the halo spin, and then the alignment starts to evolve with the interaction.

The frequent occurrence of prograde-polar interactions for halos residing in the filament is associated with the spin-flip phenomenon in the filament.
Halos that formed with the local filament initially have a spin parallel to the filamentary spine, but soon face the cosmic flow along the filament \citep{Codi15}.
Neighbors following the cosmic flow are candidates for misaligned interactions, and gradually affect the magnitude and direction of the halo spin \citep{Bett12,Bett16}.
The interactions incline the halo spin toward a perpendicular direction with respect to the initial angular momentum by converting the neighbor's orbital angular momentum into the target halo's spin \citep[e.g.,][]{Aube04,Bail05}.
In this vein, our Galaxy seems to be currently undergoing spin-flip.
\citet{Libe15} revealed that the Galaxy resides in the local filament and has a plane of satellites perpendicular to its disk plane.
If our Galaxy has a spin parallel to the filament, the satellite galaxies are in the process of infall along the local filament, and in turn, their orbital angular momenta are perpendicular to the Galactic rotation axis.
With the combination of the observational findings and theoretical hypotheses, we infer that the satellite galaxies will induce the spin-flip of our Galaxy.
The formation of the initial SOA is thus strongly linked to the cosmic flow along the LSS.

Based on the notion of the spin-flip, we propose a scenario about how the spin direction of galaxies around the filament evolves.
In the outskirts of a filament, the spin direction tends to be random with respect to the filament regardless of the mass.
After falling onto the filamentary line, galaxies gravitationally interact with neighbors coming along the filamentary line \citep{Wang05,Wang14,Libe14,Shi15,WK17} and their spin directions change into being perpendicular to the filament.
This process imprints prograde-polar alignment in the spin--orbit angle distribution.
On the other hand, the degree of spin-flip in the filament depends on the halo mass (see the right panel of Figure \ref{fig:sfalign}): the more massive the halo is, the lower the median value of $\cos\theta_{\rm SF}$ is.
This seems due to the difference in infall history: more massive galaxies enter the filament earlier than less massive ones.
The galaxy spin-flip in our scenario has been observed in both simulations \citep[e.g.,][]{Kral20} and observations using integral field spectroscopy \citep[e.g.,][]{Welk20}.
We note that the galaxy spin is more vulnerable than the halo spin due to hydrodynamic effects (such as disk instability and stellar feedback) as well as gravitational effects. 
The SOA for galaxies is stronger than that for dark matter halos \citepalias{Moon21} while the intrinsic alignment for galaxies is weaker than that for dark matter halos \citep{Chis17,Codi18}.
Such differences coincide with the galaxy--halo misalignment induced by various physical causes \citep{Bail05,Vell15a,Shao16,Chis17}.
The galaxy spin, however, has followed and will be aligned with the halo spin after all \citep{Okab20}.
In addition, we used the spin direction measured only in the inner part of the halo because the inner halo spin better represents the galaxy spin \citep{Bail05b,Bett10,Shao16}.
Hence, it is legitimate to address the SOA phenomenon using dark matter halos, and the SOA of halos is crucial to understanding the evolution of the galaxy spin.

\subsection{Angular Momentum Transfer to the Core Radius} \label{sec:angmomtransfer}

We have discussed the formation and evolution of dark matter halos' SOA in view of the tidal effect of both interacting neighbors and the anisotropic LSS.
The halo spin used in this paper is defined as the spin of the inner region of the halo (see Section \ref{sec:soangmeasure}).
Our definition of the halo spin enables us (a) to avoid contamination by the false member assignment of the halo finding algorithm by which part of member particles of the interacting neighbor are technically assigned to the outskirts of the target halo and (b) to link the SOA for dark matter halos to that for galaxies.
Despite these advantages and our new findings, the question remains whether the external tidal force directly affects the inner halo region.

Figure \ref{fig:SpinRadius} shows the offset angle ($\theta_{\vect{S_{<R}}\vect{S_{\rm w}}}$) between the spin direction measured within a certain radius ($\vect{S_{<R}}$) and that measured within the virial radius ($\vect{S_{\rm w}}$).
We divide the target halo sample into three subsamples: (a) target halos without interacting neighbors having a distance smaller than the sum of the virial radii of constituent halos of the target--neighbor pair ($d_{\rm tn}<R_{\rm vir,t}+R_{\rm vir,n}$), (b) those with only distant neighbors meeting our distance criterion ($0.8R_{\rm vir,t}<d_{\rm tn}<R_{\rm vir,t}+R_{\rm vir,n}$), and (c) those with close neighbors having a distance smaller than $0.8R_{\rm vir,t}$ ($d_{\rm tn}<0.8R_{\rm vir,t}$).
In all panels, as the radius increases, the median value of $\cos\theta_{\vect{S_{<R}}\vect{S_{\rm w}}}$ reaches to 1.0---that is, $\theta_{\vect{S_{<R}}\vect{S_{\rm w}}}$ decreases to 0 deg.
In the middle panel, the median value for target halos with distant interacting neighbors is lower than that for target halos without interacting neighbors at all radii.
This indicates that interacting neighbors first disturb the outskirts of the halo and then the offset angle increases overall.
In the right panel, the median value for target halos with close neighbors is greater at $R>0.4R_{\rm vir}$ but lower at $R<0.4R_{\rm vir}$, compared to that for target halos without interacting neighbors.
As for the different trends between the inner and outer halo regions, there are two possibilities: (i) the close interacting neighbors exert the tidal torque on their target halo and transfer their orbital angular momenta to the target halo's spin and (ii) the halo finder can inaccurately assign member particles of the neighbors to members of the target during the contact interaction (i.e., overlapping the virial radius).
In this paper, we decided to avert the second possibility by using the inner halo spin.

We undoubtedly expect that the external tidal force is exerted even on the inner region of the target halo and then the inner halo spin is modified in its direction.
This is shown as the dependence of the halo SOA on the pairwise distance (Figure \ref{fig:pairdist}) and the distance from the nearest filament (Figure \ref{fig:dndf}).
\citet{Moon21} found the galaxy SOA for both their stellar and dark matter components even though the galaxy resides in the inner part of the dark matter halo.
Furthermore, in Figure \ref{fig:SpinRadius} (b), the offset angle between the spin directions in the core region is smaller for target halos with no interacting neighbor than for those with neighbors.
This implies that the inner halo spin direction will be realigned with the whole halo spin direction during the virialization of the halo \citep{Pich11}.
In other words, the tidal interactions with the neighbor and the anisotropic LSS first influence the outskirts of the halo, and then the inner part of the halo is affected.
There will be a time lag in transferring the angular momentum from the outer region to the inner region.
To prove our speculation, we will investigate the time evolution of SOA in a forthcoming paper in this series.

\section{Summary and Conclusion} \label{sec:sum}

In an attempt to understand how galactic angular momenta develop, we have analyzed the spin--orbit alignment (SOA) between the spin of a target halo and the orbital angular momentum of its neighbor using cosmological dark matter simulations.
We selected a target--neighbor pair sample with a target halo mass of $10^{10.8}-10^{13.0}\hMsun$, a mass ratio of $1/10-10$, and a pairwise distance ranging from $0.8R_{\rm vir,t}$ (i.e., a criterion to prevent an artificial SOA by false member particle assignment) to the sum of the virial radii of the paired halos ($R_{\rm vir,t}+R_{\rm vir,n}$).
To quantify the SOA, we measured the spin--orbit angle between the halo spin vector and the neighbor's orbital angular momentum vector.
For a more detailed interpretation, we examined the dependence of SOA on both parameters related to interactions with neighbors (i.e., the halo mass, mass ratio, pairwise distance, and interaction type) and those related to the LSS (i.e., the large-scale density, the distances from the nearest filament and node, and the degree of spin--filament alignment).
Our main results are summarized as follows:

\begin{enumerate}[(1)]

\item
The spin of a target halo is well aligned with the orbital angular momentum of its neighbor whose mass ranges from one-tenth of the target's mass to 10 times this mass (Figure \ref{fig:tsoasig}).
For all of the target--neighbor pair sample, the amplitude of the alignment ($n(\cossl)$) monotonically increases from 0.92 at $\cossl<-0.75$ to 1.11 at $\cossl>0.75$ (1.0 for random alignment), and the prograde fraction ($f_{\rm prog}$) is 52.7\,\%\,$\pm$\,0.2\,\% (50\,\% for random alignment).
The significant prograde alignment comes from the tidal effects by both the interacting neighbor and the anisotropic LSS.

\item
SOA gets stronger by interaction with neighbors, in view of the higher prograde fraction (up to 56\,\%) for less massive halos with close, merging neighbors compared to that for more massive halos with distant, flybying neighbors (Figures \ref{fig:massenv}, \ref{fig:pairdist} and \ref{fig:soamgfb}).
This is attributed to two facts: (a) a less massive halo is more susceptible to tidal interaction, and (b) a close, merging neighbor has affected the halo spin direction for a longer duration of the interaction.
The neighbor's orbital angular momentum is thus gradually turned into the internal angular momentum of the target halo.

\item

Halos closer to the filament and with merging (flybying) neighbors have a stronger (weaker) SOA than those farther from the filament (Figure \ref{fig:dndf}) while the SOA shows no dependence on the large-scale density.
More interestingly, halos with the spin parallel to the filament experience merger interactions most frequently with neighbors on the prograde-polar orbit ($\cossl\simeq0.4$) and flyby encounters most frequently with neighbors on the prograde orbit ($\cossl\simeq1.0$) (Figure \ref{fig:soasf}).
Among the interactions, the prograde-polar merger interaction turns the direction of the halo spin from being parallel to the filament into being perpendicular, i.e., the spin-flip phenomenon.

\end{enumerate}

Our results lead us to conclude that the SOA originates from the tidal effects by the anisotropic matter distribution such as that of the filament and then grows by tidal interactions with neighbors.
We conjecture that the contributions of the two physical causes (i.e., neighbor-interaction and the LSS) to SOA are more or less comparable.
It is, however, hard to quantitatively compare between the contributions because a galaxy comes along the cosmic flow generated by the LSS and at the same time the galaxy, as a neighbor, tidally affects its target halo.
Hence, the two physical origins are linked to each other.
To disentangle them, we will explore the evolution of the SOA by tracing the spin direction and magnitude during the interaction with neighbors depending on the trajectories with respect to the filament.

\acknowledgments

S.-J.Y. acknowledges support from the Mid-career Researcher Program (No. 2019R1A2C3006242) and the SRC Program (the Center for Galaxy Evolution Research; No. 2017R1A5A1070354) through the National Research Foundation of Korea. This work was supported by the National Supercomputing Center with supercomputing resources including technical support (KSC-2015-C3-010 and KSC-2020-CRE-0115).

\bibliography{ref}

\begin{thebibliography}{82}
\expandafter\ifx\csname natexlab\endcsname\relax\def\natexlab#1{#1}\fi

\bibitem[{{Agustsson} \& {Brainerd}(2010)}]{Agus10}
{Agustsson}, I., \& {Brainerd}, T.~G. 2010, \apj, 709, 1321

\bibitem[{{An} {et~al.}(2019){An}, {Kim}, {Moon}, \& {Yoon}}]{An19}
{An}, S.-H., {Kim}, J., {Moon}, J.-S., \& {Yoon}, S.-J. 2019, \apj, 887, 59

\bibitem[{{Arag{\'o}n-Calvo} {et~al.}(2007){Arag{\'o}n-Calvo}, {van de
  Weygaert}, {Jones}, \& {van der Hulst}}]{Arag07}
{Arag{\'o}n-Calvo}, M.~A., {van de Weygaert}, R., {Jones}, B.~J.~T., \& {van
  der Hulst}, J.~M. 2007, \apjl, 655, L5

\bibitem[{{Aubert} {et~al.}(2004){Aubert}, {Pichon}, \& {Colombi}}]{Aube04}
{Aubert}, D., {Pichon}, C., \& {Colombi}, S. 2004, \mnras, 352, 376

\bibitem[{{Bailin} \& {Steinmetz}(2005)}]{Bail05}
{Bailin}, J., \& {Steinmetz}, M. 2005, \apj, 627, 647

\bibitem[{{Bailin} {et~al.}(2005){Bailin}, {Kawata}, {Gibson}, {Steinmetz},
  {Navarro}, {Brook}, {Gill}, {Ibata}, {Knebe}, {Lewis}, \&
  {Okamoto}}]{Bail05b}
{Bailin}, J., {Kawata}, D., {Gibson}, B.~K., {et~al.} 2005, \apjl, 627, L17

\bibitem[{{Behroozi} {et~al.}(2013){Behroozi}, {Wechsler}, \& {Wu}}]{Behr13}
{Behroozi}, P.~S., {Wechsler}, R.~H., \& {Wu}, H.-Y. 2013, \apj, 762, 109

\bibitem[{{Bennett} {et~al.}(2013){Bennett}, {Larson}, {Weiland}, {Jarosik},
  {Hinshaw}, {Odegard}, {Smith}, {Hill}, {Gold}, \& {Halpern}}]{Benn13}
{Bennett}, C.~L., {Larson}, D., {Weiland}, J.~L., {et~al.} 2013, \apjs, 208, 20

\bibitem[{{Bett} {et~al.}(2007){Bett}, {Eke}, {Frenk}, {Jenkins}, {Helly}, \&
  {Navarro}}]{Bett07}
{Bett}, P., {Eke}, V., {Frenk}, C.~S., {et~al.} 2007, \mnras, 376, 215

\bibitem[{{Bett} {et~al.}(2010){Bett}, {Eke}, {Frenk}, {Jenkins}, \&
  {Okamoto}}]{Bett10}
{Bett}, P., {Eke}, V., {Frenk}, C.~S., {Jenkins}, A., \& {Okamoto}, T. 2010,
  \mnras, 404, 1137

\bibitem[{{Bett} \& {Frenk}(2012)}]{Bett12}
{Bett}, P.~E., \& {Frenk}, C.~S. 2012, \mnras, 420, 3324

\bibitem[{{Bett} \& {Frenk}(2016)}]{Bett16}
---. 2016, \mnras, 461, 1338

\bibitem[{{Binney} \& {Tremaine}(1987)}]{BT87}
{Binney}, J., \& {Tremaine}, S. 1987, {Galactic dynamics} (Princeton, NJ:
  Princeton Univ. Press)

\bibitem[{{Blue Bird} {et~al.}(2020){Blue Bird}, {Davis}, {Luber}, {van
  Gorkom}, {Wilcots}, {Pisano}, {Gim}, {Momjian}, {Fernandez}, {Hess},
  {Lucero}, {Dodson}, {Vinsen}, {Popping}, {Chung}, {Kreckel}, {van der Hulst},
  \& {Yun}}]{BB20}
{Blue Bird}, J., {Davis}, J., {Luber}, N., {et~al.} 2020, \mnras, 492, 153

\bibitem[{{Boylan-Kolchin} {et~al.}(2008){Boylan-Kolchin}, {Ma}, \&
  {Quataert}}]{Boyl08}
{Boylan-Kolchin}, M., {Ma}, C.-P., \& {Quataert}, E. 2008, \mnras, 383, 93

\bibitem[{{Brainerd}(2005)}]{Brai05}
{Brainerd}, T.~G. 2005, \apjl, 628, L101

\bibitem[{{Bryan} \& {Norman}(1998)}]{Brya98}
{Bryan}, G.~L., \& {Norman}, M.~L. 1998, \apj, 495, 80

\bibitem[{{Bullock} {et~al.}(2001){Bullock}, {Dekel}, {Kolatt}, {Kravtsov},
  {Klypin}, {Porciani}, \& {Primack}}]{Bull01}
{Bullock}, J.~S., {Dekel}, A., {Kolatt}, T.~S., {et~al.} 2001, \apj, 555, 240

\bibitem[{{Buxton} \& {Ryden}(2012)}]{Buxt12}
{Buxton}, J., \& {Ryden}, B.~S. 2012, \apj, 756, 135

\bibitem[{{Cappellari} {et~al.}(2011){Cappellari}, {Emsellem}, {Krajnovi{\'c}},
  {McDermid}, {Serra}, {Alatalo}, {Blitz}, {Bois}, {Bournaud}, {Bureau},
  {Davies}, {Davis}, {de Zeeuw}, {Khochfar}, {Kuntschner}, {Lablanche},
  {Morganti}, {Naab}, {Oosterloo}, {Sarzi}, {Scott}, {Weijmans}, \&
  {Young}}]{Cape11}
{Cappellari}, M., {Emsellem}, E., {Krajnovi{\'c}}, D., {et~al.} 2011, \mnras,
  416, 1680

\bibitem[{{Cautun} {et~al.}(2014){Cautun}, {van de Weygaert}, {Jones}, \&
  {Frenk}}]{Caut14}
{Cautun}, M., {van de Weygaert}, R., {Jones}, B. J.~T., \& {Frenk}, C.~S. 2014,
  \mnras, 441, 2923

\bibitem[{{Cervantes-Sodi} {et~al.}(2010){Cervantes-Sodi}, {Hernandez}, \&
  {Park}}]{Cerv10}
{Cervantes-Sodi}, B., {Hernandez}, X., \& {Park}, C. 2010, \mnras, 402, 1807

\bibitem[{{Chisari} {et~al.}(2017){Chisari}, {Koukoufilippas}, {Jindal},
  {Peirani}, {Beckmann}, {Codis}, {Devriendt}, {Miller}, {Dubois}, {Laigle},
  {Slyz}, \& {Pichon}}]{Chis17}
{Chisari}, N.~E., {Koukoufilippas}, N., {Jindal}, A., {et~al.} 2017, \mnras,
  472, 1163

\bibitem[{{Codis} {et~al.}(2018){Codis}, {Jindal}, {Chisari}, {Vibert},
  {Dubois}, {Pichon}, \& {Devriendt}}]{Codi18}
{Codis}, S., {Jindal}, A., {Chisari}, N.~E., {et~al.} 2018, \mnras, 481, 4753

\bibitem[{{Codis} {et~al.}(2012){Codis}, {Pichon}, {Devriendt}, {Slyz},
  {Pogosyan}, {Dubois}, \& {Sousbie}}]{Codi12}
{Codis}, S., {Pichon}, C., {Devriendt}, J., {et~al.} 2012, \mnras, 427, 3320

\bibitem[{{Codis} {et~al.}(2015){Codis}, {Pichon}, \& {Pogosyan}}]{Codi15}
{Codis}, S., {Pichon}, C., \& {Pogosyan}, D. 2015, \mnras, 452, 3369

\bibitem[{{Darg} {et~al.}(2011){Darg}, {Kaviraj}, {Lintott}, {Schawinski},
  {Silk}, {Lynn}, {Bamford}, \& {Nichol}}]{Darg11}
{Darg}, D.~W., {Kaviraj}, S., {Lintott}, C.~J., {et~al.} 2011, \mnras, 416,
  1745

\bibitem[{{Dong} {et~al.}(2014){Dong}, {Lin}, {Kang}, {Ocean Wang}, {Dutton},
  \& {Macci{\`o}}}]{Dong14}
{Dong}, X.~C., {Lin}, W.~P., {Kang}, X., {et~al.} 2014, \apjl, 791, L33

\bibitem[{{Dubinski}(1992)}]{Dubi92}
{Dubinski}, J. 1992, \apj, 401, 441

\bibitem[{{Dubinski} {et~al.}(2004){Dubinski}, {Kim}, {Park}, \&
  {Humble}}]{Dubi04}
{Dubinski}, J., {Kim}, J., {Park}, C., \& {Humble}, R. 2004, \na, 9, 111

\bibitem[{{Fernando} {et~al.}(2017){Fernando}, {Arias}, {Guglielmo}, {Lewis},
  {Ibata}, \& {Power}}]{Fern17}
{Fernando}, N., {Arias}, V., {Guglielmo}, M., {et~al.} 2017, \mnras, 465, 641

\bibitem[{{Forero-Romero} {et~al.}(2014){Forero-Romero}, {Contreras}, \&
  {Padilla}}]{Fore14}
{Forero-Romero}, J.~E., {Contreras}, S., \& {Padilla}, N. 2014, \mnras, 443,
  1090

\bibitem[{{Ganeshaiah Veena} {et~al.}(2019){Ganeshaiah Veena}, {Cautun},
  {Tempel}, {van de Weygaert}, \& {Frenk}}]{Gane19}
{Ganeshaiah Veena}, P., {Cautun}, M., {Tempel}, E., {van de Weygaert}, R., \&
  {Frenk}, C.~S. 2019, Monthly Notices of the Royal Astronomical Society, 487,
  1607

\bibitem[{{Ganeshaiah Veena} {et~al.}(2018){Ganeshaiah Veena}, {Cautun}, {van
  de Weygaert}, {Tempel}, {Jones}, {Rieder}, \& {Frenk}}]{Gane18}
{Ganeshaiah Veena}, P., {Cautun}, M., {van de Weygaert}, R., {et~al.} 2018,
  Monthly Notices of the Royal Astronomical Society, 481, 414

\bibitem[{{Gnedin}(2003)}]{Gned03}
{Gnedin}, O.~Y. 2003, \apj, 582, 141

\bibitem[{{Hahn} {et~al.}(2007){Hahn}, {Carollo}, {Porciani}, \&
  {Dekel}}]{Hahn07}
{Hahn}, O., {Carollo}, C.~M., {Porciani}, C., \& {Dekel}, A. 2007, \mnras, 381,
  41

\bibitem[{{Herbert-Fort} {et~al.}(2008){Herbert-Fort}, {Zaritsky}, {Jin Kim},
  {Bailin}, \& {Taylor}}]{Herb08}
{Herbert-Fort}, S., {Zaritsky}, D., {Jin Kim}, Y., {Bailin}, J., \& {Taylor},
  J.~E. 2008, \mnras, 384, 803

\bibitem[{{Hetznecker} \& {Burkert}(2006)}]{Hetz06}
{Hetznecker}, H., \& {Burkert}, A. 2006, \mnras, 370, 1905

\bibitem[{{Hwang} \& {Park}(2010)}]{Hwan10}
{Hwang}, H.~S., \& {Park}, C. 2010, \apj, 720, 522

\bibitem[{{Jiang} {et~al.}(2008){Jiang}, {Jing}, {Faltenbacher}, {Lin}, \&
  {Li}}]{Jian08}
{Jiang}, C.~Y., {Jing}, Y.~P., {Faltenbacher}, A., {Lin}, W.~P., \& {Li}, C.
  2008, \apj, 675, 1095

\bibitem[{{Johnson} {et~al.}(2019){Johnson}, {Maller}, {Berlind}, {Sinha}, \&
  {Holley-Bockelmann}}]{John19}
{Johnson}, J.~W., {Maller}, A.~H., {Berlind}, A.~A., {Sinha}, M., \&
  {Holley-Bockelmann}, J.~K. 2019, \mnras, 486, 1156

\bibitem[{{Kang} \& {Wang}(2015)}]{Kang15}
{Kang}, X., \& {Wang}, P. 2015, \apj, 813, 6

\bibitem[{{Kim} {et~al.}(2018){Kim}, {Jeong}, {Lee}, {Lee}, {Joo}, {Kim}, \&
  {Rey}}]{KimS18}
{Kim}, S., {Jeong}, H., {Lee}, J., {et~al.} 2018, \apjl, 860, L3

\bibitem[{{Koo} \& {Lee}(2018)}]{Koo18}
{Koo}, H., \& {Lee}, J. 2018, \apj, 858, 51

\bibitem[{{Kraljic} {et~al.}(2020){Kraljic}, {Dav{\'e}}, \& {Pichon}}]{Kral20}
{Kraljic}, K., {Dav{\'e}}, R., \& {Pichon}, C. 2020, \mnras, 493, 362

\bibitem[{{Kraljic} {et~al.}(2018){Kraljic}, {Arnouts}, {Pichon}, {Laigle}, {de
  la Torre}, {Vibert}, {Cadiou}, {Dubois}, {Treyer}, {Schimd}, {Codis}, {de
  Lapparent}, {Devriendt}, {Hwang}, {Le Borgne}, {Malavasi}, {Milliard},
  {Musso}, {Pogosyan}, {Alpaslan}, {Bland-Hawthorn}, \& {Wright}}]{Kral18}
{Kraljic}, K., {Arnouts}, S., {Pichon}, C., {et~al.} 2018, \mnras, 474, 547

\bibitem[{{Krolewski} {et~al.}(2019){Krolewski}, {Ho}, {Chen}, {Chan},
  {Tenneti}, {Bizyaev}, \& {Kraljic}}]{Krol19}
{Krolewski}, A., {Ho}, S., {Chen}, Y.-C., {et~al.} 2019, \apj, 876, 52

\bibitem[{{Lacey} \& {Cole}(1993)}]{LC93}
{Lacey}, C., \& {Cole}, S. 1993, \mnras, 262, 627

\bibitem[{{Lagos} {et~al.}(2018){Lagos}, {Stevens}, {Bower}, {Davis},
  {Contreras}, {Padilla}, {Obreschkow}, {Croton}, {Trayford}, {Welker}, \&
  {Theuns}}]{Lago18}
{Lagos}, C.~d.~P., {Stevens}, A.~R.~H., {Bower}, R.~G., {et~al.} 2018, \mnras,
  473, 4956

\bibitem[{{Laigle} {et~al.}(2015){Laigle}, {Pichon}, {Codis}, {Dubois}, {Le
  Borgne}, {Pogosyan}, {Devriendt}, {Peirani}, {Prunet}, {Rouberol}, {Slyz}, \&
  {Sousbie}}]{Laig15}
{Laigle}, C., {Pichon}, C., {Codis}, S., {et~al.} 2015, \mnras, 446, 2744

\bibitem[{{Lee}(2012)}]{LeeJ12}
{Lee}, J. 2012, \apj, 751, 153

\bibitem[{{Lee} \& {Lemson}(2013)}]{LL13}
{Lee}, J., \& {Lemson}, G. 2013, \jcap, 2013, 022

\bibitem[{{Lee} {et~al.}(2020){Lee}, {Pak}, \& {Lee}}]{LeeJ20}
{Lee}, J.~H., {Pak}, M., \& {Lee}, H.-R. 2020, \apj, 893, 154

\bibitem[{{Lee} {et~al.}(2019{\natexlab{a}}){Lee}, {Pak}, {Lee}, \&
  {Song}}]{LeeJ19a}
{Lee}, J.~H., {Pak}, M., {Lee}, H.-R., \& {Song}, H. 2019{\natexlab{a}}, \apj,
  872, 78

\bibitem[{{Lee} {et~al.}(2019{\natexlab{b}}){Lee}, {Pak}, {Song}, {Lee}, {Kim},
  \& {Jeong}}]{LeeJ19b}
{Lee}, J.~H., {Pak}, M., {Song}, H., {et~al.} 2019{\natexlab{b}}, \apj, 884,
  104

\bibitem[{{L'Huillier} {et~al.}(2017){L'Huillier}, {Park}, \& {Kim}}]{LHui17}
{L'Huillier}, B., {Park}, C., \& {Kim}, J. 2017, \mnras, 466, 4875

\bibitem[{{Libeskind} {et~al.}(2005){Libeskind}, {Frenk}, {Cole}, {Helly},
  {Jenkins}, {Navarro}, \& {Power}}]{Libe05}
{Libeskind}, N.~I., {Frenk}, C.~S., {Cole}, S., {et~al.} 2005, \mnras, 363, 146

\bibitem[{{Libeskind} {et~al.}(2015){Libeskind}, {Hoffman}, {Tully},
  {Courtois}, {Pomar{\`e}de}, {Gottl{\"o}ber}, \& {Steinmetz}}]{Libe15}
{Libeskind}, N.~I., {Hoffman}, Y., {Tully}, R.~B., {et~al.} 2015, Monthly
  Notices of the Royal Astronomical Society, 452, 1052

\bibitem[{{Libeskind} {et~al.}(2014){Libeskind}, {Knebe}, {Hoffman}, \&
  {Gottl{\"o}ber}}]{Libe14}
{Libeskind}, N.~I., {Knebe}, A., {Hoffman}, Y., \& {Gottl{\"o}ber}, S. 2014,
  \mnras, 443, 1274

\bibitem[{{Libeskind} {et~al.}(2011){Libeskind}, {Knebe}, {Hoffman},
  {Gottl{\"o}ber}, {Yepes}, \& {Steinmetz}}]{Libe11}
{Libeskind}, N.~I., {Knebe}, A., {Hoffman}, Y., {et~al.} 2011, \mnras, 411,
  1525

\bibitem[{{Marinacci} {et~al.}(2018){Marinacci}, {Vogelsberger}, {Pakmor},
  {Torrey}, {Springel}, {Hernquist}, {Nelson}, {Weinberger}, {Pillepich},
  {Naiman}, \& {Genel}}]{Mari18}
{Marinacci}, F., {Vogelsberger}, M., {Pakmor}, R., {et~al.} 2018, \mnras, 480,
  5113

\bibitem[{{Mesa} {et~al.}(2014){Mesa}, {Duplancic}, {Alonso}, {Coldwell}, \&
  {Lambas}}]{Mesa14}
{Mesa}, V., {Duplancic}, F., {Alonso}, S., {Coldwell}, G., \& {Lambas}, D.~G.
  2014, \mnras, 438, 1784

\bibitem[{{Mesa} {et~al.}(2018){Mesa}, {Duplancic}, {Alonso}, {Mu{\~n}oz
  Jofr{\'e}}, {Coldwell}, \& {Lambas}}]{Mesa18}
{Mesa}, V., {Duplancic}, F., {Alonso}, S., {et~al.} 2018, \aap, 619, A24

\bibitem[{{Moon} {et~al.}(2019){Moon}, {An}, \& {Yoon}}]{Moon19}
{Moon}, J.-S., {An}, S.-H., \& {Yoon}, S.-J. 2019, \apj, 882, 14

\bibitem[{{Moon} {et~al.}(2021){Moon}, {An}, \& {Yoon}}]{Moon21}
---. 2021, \apj, 909, 34

\bibitem[{{Muldrew} {et~al.}(2011){Muldrew}, {Pearce}, \& {Power}}]{Muld11}
{Muldrew}, S.~I., {Pearce}, F.~R., \& {Power}, C. 2011, \mnras, 410, 2617

\bibitem[{{Musso} {et~al.}(2018){Musso}, {Cadiou}, {Pichon}, {Codis},
  {Kraljic}, \& {Dubois}}]{Muss18}
{Musso}, M., {Cadiou}, C., {Pichon}, C., {et~al.} 2018, \mnras, 476, 4877

\bibitem[{{Naiman} {et~al.}(2018){Naiman}, {Pillepich}, {Springel},
  {Ramirez-Ruiz}, {Torrey}, {Vogelsberger}, {Pakmor}, {Nelson}, {Marinacci},
  {Hernquist}, {Weinberger}, \& {Genel}}]{Naim18}
{Naiman}, J.~P., {Pillepich}, A., {Springel}, V., {et~al.} 2018, \mnras, 477,
  1206

\bibitem[{{Nelson} {et~al.}(2018){Nelson}, {Pillepich}, {Springel},
  {Weinberger}, {Hernquist}, {Pakmor}, {Genel}, {Torrey}, {Vogelsberger},
  {Kauffmann}, {Marinacci}, \& {Naiman}}]{Nels18}
{Nelson}, D., {Pillepich}, A., {Springel}, V., {et~al.} 2018, \mnras, 475, 624

\bibitem[{{Okabe} {et~al.}(2020){Okabe}, {Nishimichi}, {Oguri}, {Peirani},
  {Kitayama}, {Sasaki}, {Suto}, {Pichon}, \& {Dubois}}]{Okab20}
{Okabe}, T., {Nishimichi}, T., {Oguri}, M., {et~al.} 2020, \mnras, 491, 2268

\bibitem[{{Peebles}(1969)}]{Peeb69}
{Peebles}, P.~J.~E. 1969, \apj, 155, 393

\bibitem[{{Peirani} {et~al.}(2004){Peirani}, {Mohayaee}, \& {de Freitas
  Pacheco}}]{Peir04}
{Peirani}, S., {Mohayaee}, R., \& {de Freitas Pacheco}, J.~A. 2004, \mnras,
  348, 921

\bibitem[{{Peng} \& {Renzini}(2020)}]{PR20}
{Peng}, Y.-j., \& {Renzini}, A. 2020, \mnras, 491, L51

\bibitem[{{Pichon} {et~al.}(2011){Pichon}, {Pogosyan}, {Kimm}, {Slyz},
  {Devriendt}, \& {Dubois}}]{Pich11}
{Pichon}, C., {Pogosyan}, D., {Kimm}, T., {et~al.} 2011, \mnras, 418, 2493

\bibitem[{{Pillepich} {et~al.}(2018){Pillepich}, {Nelson}, {Hernquist},
  {Springel}, {Pakmor}, {Torrey}, {Weinberger}, {Genel}, {Naiman}, {Marinacci},
  \& {Vogelsberger}}]{Pill18}
{Pillepich}, A., {Nelson}, D., {Hernquist}, L., {et~al.} 2018, \mnras, 475, 648

\bibitem[{{Porciani} {et~al.}(2002){Porciani}, {Dekel}, \& {Hoffman}}]{Porc02}
{Porciani}, C., {Dekel}, A., \& {Hoffman}, Y. 2002, \mnras, 332, 325

\bibitem[{{Rodriguez-Gomez} {et~al.}(2017){Rodriguez-Gomez}, {Sales}, {Genel},
  {Pillepich}, {Zjupa}, {Nelson}, {Griffen}, {Torrey}, {Snyder},
  {Vogelsberger}, {Springel}, {Ma}, \& {Hernquist}}]{Rodr17}
{Rodriguez-Gomez}, V., {Sales}, L.~V., {Genel}, S., {et~al.} 2017, \mnras, 467,
  3083

\bibitem[{{Sales} {et~al.}(2012){Sales}, {Navarro}, {Theuns}, {Schaye},
  {White}, {Frenk}, {Crain}, \& {Dalla Vecchia}}]{Sale12}
{Sales}, L.~V., {Navarro}, J.~F., {Theuns}, T., {et~al.} 2012, \mnras, 423,
  1544

\bibitem[{{Shao} {et~al.}(2016){Shao}, {Cautun}, {Frenk}, {Gao}, {Crain},
  {Schaller}, {Schaye}, \& {Theuns}}]{Shao16}
{Shao}, S., {Cautun}, M., {Frenk}, C.~S., {et~al.} 2016, \mnras, 460, 3772

\bibitem[{{Shi} {et~al.}(2015){Shi}, {Wang}, \& {Mo}}]{Shi15}
{Shi}, J., {Wang}, H., \& {Mo}, H.~J. 2015, \apj, 807, 37

\bibitem[{{Sousbie}(2011)}]{Sous11a}
{Sousbie}, T. 2011, \mnras, 414, 350

\bibitem[{{Sousbie} {et~al.}(2011){Sousbie}, {Pichon}, \& {Kawahara}}]{Sous11b}
{Sousbie}, T., {Pichon}, C., \& {Kawahara}, H. 2011, \mnras, 414, 384

\bibitem[{{Springel} {et~al.}(2018){Springel}, {Pakmor}, {Pillepich},
  {Weinberger}, {Nelson}, {Hernquist}, {Vogelsberger}, {Genel}, {Torrey},
  {Marinacci}, \& {Naiman}}]{Spri18}
{Springel}, V., {Pakmor}, R., {Pillepich}, A., {et~al.} 2018, \mnras, 475, 676

\bibitem[{{Stewart} {et~al.}(2013){Stewart}, {Brooks}, {Bullock}, {Maller},
  {Diemand}, {Wadsley}, \& {Moustakas}}]{Stew13}
{Stewart}, K.~R., {Brooks}, A.~M., {Bullock}, J.~S., {et~al.} 2013, \apj, 769,
  74

\bibitem[{{Tempel} {et~al.}(2015){Tempel}, {Guo}, {Kipper}, \&
  {Libeskind}}]{Temp15}
{Tempel}, E., {Guo}, Q., {Kipper}, R., \& {Libeskind}, N.~I. 2015, \mnras, 450,
  2727

\bibitem[{{Tempel} \& {Libeskind}(2013)}]{Temp13}
{Tempel}, E., \& {Libeskind}, N.~I. 2013, \apjl, 775, L42

\bibitem[{{Tempel} \& {Tamm}(2015)}]{TT15}
{Tempel}, E., \& {Tamm}, A. 2015, \aap, 576, L5

\bibitem[{{Velliscig} {et~al.}(2015{\natexlab{a}}){Velliscig}, {Cacciato},
  {Schaye}, {Hoekstra}, {Bower}, {Crain}, {van Daalen}, {Furlong}, {McCarthy},
  {Schaller}, \& {Theuns}}]{Vell15b}
{Velliscig}, M., {Cacciato}, M., {Schaye}, J., {et~al.} 2015{\natexlab{a}},
  \mnras, 454, 3328

\bibitem[{{Velliscig} {et~al.}(2015{\natexlab{b}}){Velliscig}, {Cacciato},
  {Schaye}, {Crain}, {Bower}, {van Daalen}, {Dalla Vecchia}, {Frenk},
  {Furlong}, {McCarthy}, {Schaller}, \& {Theuns}}]{Vell15a}
---. 2015{\natexlab{b}}, \mnras, 453, 721

\bibitem[{{Vitvitska} {et~al.}(2002){Vitvitska}, {Klypin}, {Kravtsov},
  {Wechsler}, {Primack}, \& {Bullock}}]{Vitv02}
{Vitvitska}, M., {Klypin}, A.~A., {Kravtsov}, A.~V., {et~al.} 2002, \apj, 581,
  799

\bibitem[{{Wang} {et~al.}(2011){Wang}, {Mo}, {Jing}, {Yang}, \&
  {Wang}}]{Wang11}
{Wang}, H., {Mo}, H.~J., {Jing}, Y.~P., {Yang}, X., \& {Wang}, Y. 2011, \mnras,
  413, 1973

\bibitem[{{Wang} {et~al.}(2005){Wang}, {Jing}, {Mao}, \& {Kang}}]{Wang05}
{Wang}, H.~Y., {Jing}, Y.~P., {Mao}, S., \& {Kang}, X. 2005, \mnras, 364, 424

\bibitem[{{Wang} {et~al.}(2018{\natexlab{a}}){Wang}, {Guo}, {Kang}, \&
  {Libeskind}}]{Wang18spin}
{Wang}, P., {Guo}, Q., {Kang}, X., \& {Libeskind}, N.~I. 2018{\natexlab{a}},
  \apj, 866, 138

\bibitem[{{Wang} \& {Kang}(2017)}]{WK17}
{Wang}, P., \& {Kang}, X. 2017, \mnras, 468, L123

\bibitem[{{Wang} \& {Kang}(2018)}]{WK18}
---. 2018, \mnras, 473, 1562

\bibitem[{{Wang} {et~al.}(2018{\natexlab{b}}){Wang}, {Luo}, {Kang},
  {Libeskind}, {Wang}, {Zhang}, {Tempel}, \& {Guo}}]{Wang18sca}
{Wang}, P., {Luo}, Y., {Kang}, X., {et~al.} 2018{\natexlab{b}}, \apj, 859, 115

\bibitem[{{Wang} {et~al.}(2010){Wang}, {Park}, {Hwang}, \& {Chen}}]{Wang10}
{Wang}, Y., {Park}, C., {Hwang}, H.~S., \& {Chen}, X. 2010, \apj, 718, 762

\bibitem[{{Wang} {et~al.}(2014){Wang}, {Lin}, {Kang}, {Dutton}, {Yu}, \&
  {Macci{\`o}}}]{Wang14}
{Wang}, Y.~O., {Lin}, W.~P., {Kang}, X., {et~al.} 2014, \apj, 786, 8

\bibitem[{{Warnick} \& {Knebe}(2006)}]{Warn06}
{Warnick}, K., \& {Knebe}, A. 2006, \mnras, 369, 1253

\bibitem[{{Welker} {et~al.}(2018){Welker}, {Dubois}, {Pichon}, {Devriendt}, \&
  {Chisari}}]{Welk18}
{Welker}, C., {Dubois}, Y., {Pichon}, C., {Devriendt}, J., \& {Chisari}, N.~E.
  2018, \aap, 613, A4

\bibitem[{{Welker} {et~al.}(2017){Welker}, {Power}, {Pichon}, {Dubois},
  {Devriendt}, \& {Codis}}]{Welk17}
{Welker}, C., {Power}, C., {Pichon}, C., {et~al.} 2017, arXiv e-prints,
  arXiv:1712.07818

\bibitem[{{Welker} {et~al.}(2020){Welker}, {Bland-Hawthorn}, {Van de Sande},
  {Lagos}, {Elahi}, {Obreschkow}, {Bryant}, {Pichon}, {Cortese}, {Richards},
  {Croom}, {Goodwin}, {Lawrence}, {Sweet}, {Lopez-Sanchez}, {Medling}, {Owers},
  {Dubois}, \& {Devriendt}}]{Welk20}
{Welker}, C., {Bland-Hawthorn}, J., {Van de Sande}, J., {et~al.} 2020, \mnras,
  491, 2864

\bibitem[{{White}(1984)}]{Whit84}
{White}, S.~D.~M. 1984, \apj, 286, 38

\bibitem[{{Yang} {et~al.}(2006){Yang}, {van den Bosch}, {Mo}, {Mao}, {Kang},
  {Weinmann}, {Guo}, \& {Jing}}]{Yang06}
{Yang}, X., {van den Bosch}, F.~C., {Mo}, H.~J., {et~al.} 2006, \mnras, 369,
  1293

\bibitem[{{York} {et~al.}(2000){York}, {Adelman}, {Anderson}, {Anderson},
  {Annis}, {Bahcall}, {Bakken}, {Barkhouser}, {Bastian}, \& {Berman}}]{York00}
{York}, D.~G., {Adelman}, J., {Anderson}, John~E., J., {et~al.} 2000, \aj, 120,
  1579

\bibitem[{{Zavala} {et~al.}(2016){Zavala}, {Frenk}, {Bower}, {Schaye},
  {Theuns}, {Crain}, {Trayford}, {Schaller}, \& {Furlong}}]{Zava16}
{Zavala}, J., {Frenk}, C.~S., {Bower}, R., {et~al.} 2016, \mnras, 460, 4466

\bibitem[{{Zhang} {et~al.}(2015){Zhang}, {Yang}, {Wang}, {Wang}, {Luo}, {Mo},
  \& {van den Bosch}}]{Zhan15}
{Zhang}, Y., {Yang}, X., {Wang}, H., {et~al.} 2015, \apj, 798, 17

\bibitem[{{Zjupa} \& {Springel}(2017)}]{Zjup17}
{Zjupa}, J., \& {Springel}, V. 2017, \mnras, 466, 1625

\end{thebibliography}
\bibliographystyle{aasjournal}

\end{document}